\documentclass[a4paper,11pt]{article}
\pdfoutput=1 

\usepackage{jcappub} 
\usepackage{orcidlink}
\usepackage{bbold}
\usepackage{tikz}
\usepackage{subcaption}
\usepackage{float}
\usepackage{appendix}
\usepackage{multirow}
\usepackage{enumitem}
\usepackage{amsmath}
\usepackage{natbib}
\newcommand\blfootnote[1]{%
  \begingroup
  \renewcommand\thefootnote{}\footnote{#1}%
  \addtocounter{footnote}{-1}%
  \endgroup
}

\title{Bibliography management: \texttt{natbib} package}
\author[a]{W.L. Matthewson\orcidlink{0000-0001-6957-772X}}
\author[a]{A. Shafieloo\orcidlink{0000-0001-6815-0337}}
\date{ }

\definecolor{DarkGreen}{rgb}{0,0.6,0}
\newcommand{\wlm}[1]{\textcolor{black}{#1}}
\newcommand{\wlmn}[1]{\textcolor{black}{#1}}
\newcommand{\be}{\begin{equation}}
\newcommand{\ee}{\end{equation}}
\newcommand{\ba}{\begin{align}}
\newcommand{\ea}{\end{align}}
\renewcommand{\eqref}[1]{Eq. (\ref{#1})}
\title{Star-Crossed Labours\blfootnote{
\textit{${}$\\Three datasets, alike in dignity;\\
In fair relief th'exploding stars we seat,\\
Cross'd from concordance, break degen'racy,\\
Decide the fate our toil the which shall meet.
}${ }$\\
- Adapted from the Prologue to \textit{Romeo and Juliet} by William Shakespeare.
}: Checking Consistency Between Current Supernovae Compilations}

\date{April 2024}

\begin{document}
\affiliation[a]{Korea Astronomy and Space Science Institute, 776, Daedeokdae-ro, Yuseong-gu, Daejeon 34055, Republic of Korea}
\emailAdd{willmatt4th@kasi.re.kr, shafieloo@kasi.re.kr}
\begin{abstract}
{We make use of model-independent statistical methods to assess the consistency of three different supernova compilations: Union3, Pantheon+ and DES 5-year supernovae. We expand the available model space of each, using Crossing Statistics, and test the compatibility of each dataset, against the other two. This is done using (I) a Flat $\Lambda$CDM fitting to, and (II) Iterative Smoothing from, one particular dataset, and determining the level of deformation by required to fit the other two. This allows us to test the mutual consistency of the datasets both within the standard model and in the case of some extended model, motivated by features present in a particular dataset. We find that, in both these cases, the data are only consistent with the point in the parameter space corresponding to zero deformation \wlm{(of I and II)}, at around a $2\sigma$ level, with the DES compilation showing the largest disagreement. However, all three datasets are still found to be consistent to within $1-2\sigma$ \wlm{with each other}.}
\end{abstract}
\maketitle

\section{Introduction}

The current best explanation of our Universe is the Flat $\Lambda$CDM model of cosmology. The so-called concordance model has enjoyed much success in describing many different aspects of the physical phenomena observed to date~\cite{SupernovaSearchTeam:1998fmf,SupernovaCosmologyProject:1998vns,Planck1:2020,Alam_2021,Zhao_2022}. As widely-accepted as it is, as a good descriptive model for the Universe, there are still several open questions and tensions to resolve. \cite{RevModPhys.61.1,Bull:2015stt,Perivolaropoulos:2021jda}

The exact nature of dark energy and dark matter are currently unknown; indeed, even their behaviour and evolution are the subject of ongoing study. Recently, the Dark Energy Spectroscopic Instrument (DESI) released data showing evidence for phantom dark energy \cite{desicollaboration2024desi}. The question of whether the cosmic acceleration is even due to dark energy has been raised, and does not as yet have a clear answer~\cite{Ostriker:1995su,Sahni:1999gb,Peebles:2002gy,Copeland:2006wr}.

Fortunately, in the current ``era of precision cosmology'', many surveys have as their goal the elucidation of these phenomena, and have been designed to provide further and better data with which they may be studied. \cite{DES:2017myr,DES:2017qwj,eBOSS:2020yzd,Heymans:2020gsg,DES:2021wwk,DESI2024.I.DR1}.

Addressing these issues using data, in particular on large scales and in a statistical manner, has a strong potential to aid in resolving such questions. This work will focus on the constraining power of data from supernovae specifically. Through the light curves of these so-called ``standardisable candles'', it is possible to gain access to the expansion history of the universe via the luminosity distance $D_L$, and in turn constrain certain parameters of cosmological models, such as $\Omega_m$. This is indispensable when hoping to form a clear picture of how the Universe has evolved and to gain an understanding of the underlying processes at work within it~\cite{SupernovaSearchTeam:1998fmf,SupernovaCosmologyProject:1998vns}.

Currently, another key problem within cosmology is the existence of various ``tensions'' in the estimation of cosmological model parameters. A statistical tension is a disagreement in the values of a particular model's parameters when fit to two different datasets. The significance of a given tension may be measured by considering the mutual separation of the maximal likelihood points in terms of the confidence intervals for each dataset, within the parameter space of a particular model. In general, the existence of a tension indicates that there is either a systematic in one or both of the datasets that it poorly understood, or that the model is not compatible with the data. In the case of the latter, we would need to consider extensions to the model that allow for the explanation of both datasets simultaneously. If datasets are in tension, it is not statistically sound practice to obtain constraints from the combination of the datasets: the tension in parameter values can lead to an artificial strengthening of the joint constraints, while biasing the best fit parameter values. 

In the case of supernovae data, we are not aiming to combine the datasets; indeed, given the cases of large overlap in data~\cite{desicollaboration2024desi} this would not be advisable. However, since the data are all taken from observations in the Universe, they should be mutually consistent. If we assume that the underlying model of the Universe is known (for example, the concordance model of a flat $\Lambda$CDM universe), then our observations should constitute consistent realisations of that model, to within statistical fluctuations. Consider the best fit parameter values for the model. While these should at least be without tension (assuming that the chosen model is correct), this is not a sufficient condition to conclude that the different observations are consistent. One might imagine the case where, within one fixed model, two different datasets might obtain similar best fit parameter values, even if mutually inconsistent. This simply because each dataset might be inconsistent with the other in a way to which the model is not sensitive. 
The most natural model to choose as a starting point is the concordance flat $\Lambda$CDM. In addition to this, the current open questions in cosmology suggest that we should also entertain the possibility of models beyond the standard concordance model, which might better explain the data and perhaps be sensitive to different inconsistencies between the datasets.

Recently, the possibility of systematics in the DES supernovae data being erroneously interpreted as evidence for evolving dark energy has also been raised \cite{efstathiou2024evolvingdarkenergysupernovae}. A test of consistency between available supernova datasets thus serves as a valuable cross-check as to whether the hints of behaviour outside of the concordance model originate simply as a result of the particular treatment of a dataset, or whether they might be consistent across the available observations.

In this work we aim to test the mutual consistency of various supernova datasets in a model-independent manner, where possible, making use of the Bayesian implementation of Crossing Statistics~\cite{Shafieloo:2010xm,Shafieloo:2012jb,2012JCAP...08..002S} to probe the freedoms allowed by the data, as well as the Iterative Smoothing method~\cite{Shafieloo:2005nd,Appleby_2014,Koo:2020wro} to obtain data-driven fits.
The goal is not to declare a preferred or ``best'' dataset, but rather to place them all on equal footing and draw conclusions about their mutual consistency.

The rest of this work is laid out as follows: In Section \ref{meth} we introduce the methodology used in the analysis, including a description of Crossing Statistics, Iterative Smoothing and how they will be used to judge consistency of the data, as well as a description of the data themselves. This is followed by Section \ref{temps} where we present both the modelled and model-independent ``template functions'' that will be used as a basis for the consistency analysis. We then discuss the results we obtain, in Section \ref{DnR}, before concluding in Section \ref{conc}.

\section{Methodology}
\label{meth}
In this analysis, we consider data from supernovae observations.  For a brief, yet comprehensive history of this observable, see~\cite{SupernovaSearchTeam:1998fmf}. These data are useful for cosmological applications, as supernovae can provide a distance measure for a particular redshift. Supernovae are called standardisable candles since their light curves may be fitted to a template\footnote{Not to be confused with our nomenclature of ``template function'' in the context of Crossing Statistics, see Section~\ref{Xing}.} in a consistent manner to determine their intrinsic luminosity. Thus, they give us access to the luminosity distance at the redshift measured from the host galaxy. The actual observable is the \wlm{magnitude, which may be related to cosmology by the} ``distance modulus'':
\be
\mu(z) \equiv m-M_B = 5\log_{10}\left(\frac{D_L(z)}{10{\rm pc}}\right)\,,
\label{mudef}
\ee
where $m$ is the effective (after standardisation) apparent magnitude, and $M_B$ the B-band absolute magnitude, of the standardised supernova light curve.
Although we cannot use supernovae to directly constrain $H_0$ (due to its degeneracy with $M_B$), \wlm{we can constrain the combination and, in doing so, gain access to the expansion history. For the purposes of this analysis we report the value of $M_B$ obtained when taking $h \equiv \frac{H_0}{100 km/s/Mpc} = 0.6766$, from Planck 2018 results~\cite{Planck1:2020}. With this combination fit, we may use \eqref{mudef} to} constrain parameters relating to the background expansion: $\Omega_m$, $\Omega_\Lambda$, etc. depending on the chosen cosmological model.

Currently there are various supernovae compilation datasets available, which combine supernovae from a range of observations. See Section~\ref{data} for the particulars of the datasets used in this analysis. We are interested in the mutual consistency of these datasets, following a methodology similar to~\cite{Hazra:2013oqa}, using Crossing Statistics and Iterative Smoothing, as explained in the sections to follow.

\subsection{Crossing Statistics}
\label{Xing}
The inspiration for Crossing Statistics, as first proposed in~\cite{Shafieloo:2010xm}, comes from considering two curves that model a particular observed quantity. Differences in the functional forms of these curves could lead to them `crossing' each other over the range of the independent variable considered. These crossings may be used to define a series of statistics of different orders, sensitive to the slight variations of the curves with respect to each other, that is a generalisation of the $\chi^2$ measure, and which allows for finer discrimination between the fit of various curves to observed data, particularly when there may be an unknown intrinsic error component. 
In this work, we are particularly interested in the Bayesian interpretation of these so-called ``Crossing Statistics". In this case, we consider one ``template function'' which is an estimate with a reasonable fit to some data. Leaving aside the precise origin of this function for the moment, we also consider a second function, which is obtained through small ($\sim2\%$) deformations around the template function, according to: 
\be
\hat q^n(x) = \hat q^{TF}(x) \sum_{i=0}^{n} C^n_i [T_i(x)]\,,
\label{crossing}
\ee
where $\hat q^{TF}(x)$ is the template function, $\hat q^n(x)$ is the deformation of the template function at $n^{th}$ order in Crossing Statistics and $T_n(x)$ is the $n^{th}$ order Chebyshev polynomial of the first kind\footnote{For example: \begin{flalign} T_0(x) &= 1\,, \nonumber\\T_1(x) &= x\,, \nonumber\\T_2(x) &= 2x^2-1\,, \nonumber\\T_3(x) &= 4x^3-3x\,, \nonumber\\&{\rm etc}.\nonumber\end{flalign}}. The circumflex is used here to emphasize that these are calculated estimates, and not the observed data themselves. We call $\sum_{i=0}^{n} C^n_i [T_i(x)]$ the ``crossing function'' and its expansion coefficients $C_i^n$ the ``crossing hyperparameters'', whose values will determine the shape of the resulting deformed function.

In the limit $n\rightarrow\infty$, the Chebyshev polynomials span the interval $[-1,1]$ and, as such, can provide an orthogonal basis for reconstructing different functional forms in a model-independent manner. The resulting deformations of the template function will have more degrees of freedom, and can be fit to the same data, to determine if there are any features which the data prefer that are not captured by the initial template function. The best fit crossing hyperparameters $C^n_m$ ($m\leq n$), found through some sampling method, will reveal how strongly the data support these deviations from the template function. In the case that the fiducial template function already sufficiently captures the features of the data, then \wlm{we would expect that the} values of the crossing hyperparameters for the $n^{th}$ order crossing function $(C^n_0, C^n_1, ..., C^n_n)$ \wlm{should be consistent with the point} $(1, 0, ..., 0)$, which signifies no effective deformation. Since this analysis involves tests for the consistency of the fiducial templates, we refer to these as the ``standard values'' of the crossing hyperparameters.

In practice, it is necessary to truncate the sum in \eqref{crossing} at finite order. There are various factors which must be taken into consideration, including physical constraints on the monotonicity of $\mu$ and the requirement that the fluctuations of the deformed function with respect to the template function be sufficiently small. See Appendix \ref{tuning} for a more detailed description and application to the specific datasets used in our analysis.

In the case of supernova observations, the quantity of interest is the distance modulus at values of redshift in the survey range $z_{min}<z<z_{max}$, and so the deformed functions are expressed as 
\be
\hat \mu^n(z) = \hat \mu^{TF}(z) \sum_{i=0}^{n} C^n_i T_i(\tilde z)\,,
\label{mucrossing}
\ee
where $\tilde z(z) \equiv 2\frac{z}{z_{max}}-1$ is defined to ensure that the redshift range is correctly mapped to the interval $[-1,1]$, relevant for the Chebyshev polynomials\footnote{In the case of multiple datasets and redshift ranges, $z_{max}$ is the overall maximum value of redshift where supernovae are observed, unless otherwise stated.}.

Since $\hat\mu^{TF}(z)$ must be related to the actual observed quantity $m$ through the parameter $M_B$ \wlm{which, while having one true underlying value, may have a different inferred value, depending on the subset of data observed}, with no direct cosmological significance, there is also a translational degree of freedom in the deformed function. This is not quite the same as the $0^{th}$ order crossing function; while the former provides an additive shift in the estimate for $\mu$, the latter signifies a `tilt' in the function. Thus, at all orders in Crossing Statistics, we allow for $M_B$ to vary.

Next, we discuss how to make an appropriate choice for the template function. One obvious choice is the best fit function, corresponding to a particular cosmological model, which will describe the data reasonably well. The limitation of this is that small deformations around this template function would depend somehow on the chosen cosmological model and fit. Another option, independent of cosmological parametrisation, is to construct a template function using iterative smoothing, as detailed below. This will allow us to compare the consistency at the level of the data, without inteference from a cosmological model. 

\subsection{Iterative Smoothing}
This method, introduced in~\cite{Shafieloo:2005nd}, is broadly a process meant to produce, from data, a function which has a progressively better $\chi^2$ fit to that data, in essence completely removing the residuals in the limits of infinite iterations and zero smoothing length. It relies on an initial guess function\footnote{In principle, given sufficient iterations, the precise choice of initial function should not matter.} that undergoes successive applications by a function that depends on the inverse covariance matrix of the data, the residuals and some weighting function, characterized by a smoothing length, over the domain of the data. \wlm{In this case, we use the $\Lambda$CDM best fit as the $0^{th}$ iteration of the smoothed function for a particular dataset.}

In the case of supernovae distance moduli, the estimate at arbitrary redshift $z$ in the next iteration may be written compactly as follows~\cite{Shafieloo:2018gin, LHuillier:2016mtc}:
\be
{\hat \mu}_{k+1}(z) = {\hat \mu}_k(z) + \frac{\boldsymbol{\delta\mu}_k^T \cdot {\rm C}^{-1} \cdot \mathbf{W}(z)}{\mathbb{1}^T\cdot {\rm C}^{-1} \cdot \mathbf{W}(z)}\,,
\ee
where the vectors all contain elements corresponding to each of the redshift values of the data, $z_i$. 

Consequently, the elements of the residual vector in the $k^{th}$ iteration are defined as
\be
(\boldsymbol{\delta\mu}_k)_i = {\hat\mu}_k(z_i) - \mu_i\,,
\ee
where ${\hat\mu}_k(z_i)$ is the $k^{th}$ iteration estimate for $\mu$ at each of the data point redshifts and $\mu_i$ are the corresponding data values $\mu_i = \mu(z_i)$.

$\mathbb{1} = (1, 1, ..., 1)^T$ is the column vector of ones, with length compatible to the inverse covariance matrix ${\rm C}^{-1}$,
and $\mathbf{W}(z)$ contains a distribution which weights each data point according to
\be
W_i(z) = \exp{\left(-\frac{\ln^2\left(\frac{1+z}{1+z_i}\right)}{2\Delta^2}\right)}\,,
\ee
regulated by the smoothing length $\Delta$.

In principle, this method will produce a set of points at the desired redshifts, independent of any particular cosmological model, which constitutes a smooth function whose characteristics are based solely on the data. However, there is a caveat: even if the function is independent of the initial guess, once convergence is reached (i.e. the number of iterations N must be chosen sufficiently high), the characteristics will still depend on the hyperparameter $\Delta$, which we call the smoothing length.
Thus, it is necessary that we choose the smoothing length $\Delta$ (and the maximum number of iterations, $N$) carefully. This is the subject of Appendix \ref{tuning}.

\subsection{Consistency}
As discussed previously, we would like to test the mutual consistency of the datasets in two scenarios: that of a model (flat $\Lambda$CDM) and the case where the test is independent of a particular cosmological model (instead there is an implied model which is perhaps an extension beyond the concordance cosmology). In the Bayesian application of Crossing Statistics, we deform a template function by crossing functions in order to provide more freedom and obtain a \wlm{lower $\chi^2$ statistic for} the data in a model-independent manner. \wlm{We choose the $\chi^2$ as the statistic of interest here, as an unbiased measure which is simple to calculate, given the standard practice and available likelihood codes for the data, even though the true connection to likelihood is not so straightforward.} This additional freedom gives the resulting deformed functions the possibility to be more sensitive to features of the data, which might reveal inconsistencies that go unnoticed when the data are simply fit to a cosmological model. 

For the first scenario then, we will use the flat $\Lambda$CDM model best fit as a template function and examine the behaviour of the crossing hyperparameters to test our first null hypothesis,
\begin{itemize}[leftmargin=6em]
\item[{\textbf{Null hypothesis I:}}] Given a (fit to the) assumed concordance cosmology, the datasets are mutually consistent. 
\end{itemize}
In the second case, we make use of an iteratively-smoothed function that fits the data more closely than the concordance cosmology, so that we may test the case where the underlying model is not constrained to standard $\Lambda$CDM.
The appropriate null hypothesis in this case is as follows, 
\begin{itemize}[leftmargin=6em]
\item[{\textbf{Null hypothesis II:}}] Given a fit to the data that improves on the flat $\Lambda$CDM best fit (thus implying some underlying, extended model), the datasets are mutually consistent. 
\end{itemize}

Both of these null hypotheses are actually twofold. On one hand, we hypothesise that all the datasets are consistent with the individual representation (\textit{undeformed} template function) of each dataset, under some underlying model (be it $\Lambda$CDM or not); on the other hand, we also hypothesize that there exists a volume within the implied hyperparameter space of a particular representation, where all three datasets are consistent (i.e. could all be produced as realisations from particular deformations of the underlying model). By testing each of these parts of the null hypotheses, we will be able to provide a determination of the mutual consistency of the chosen datasets\wlm{; we will thus consider the level of consistency by examining the $\sigma$ separations, in the extended parameter space, between the undeformed template function and each dataset's posterior.}
\wlm{The statistical interpretation is as follows: If two dataset contours overlap at, for example, $2-\sigma$, it means that there exists a region in the functional space (corresponding to a possible universe expansion history) which, given the corresponding model, has a $68-95\%$ probability of realising \textit{both} datasets. The threshold number of $\sigma$ is somewhat arbitray, and we leave it to the reader to impose, based on the required level of stringency.}

\subsection{Data}
\label{data}
We make use of supernova data from three survey compilations:
\begin{itemize}
\item{
``Pantheon+'' - consisting of 1550 distinct Type Ia supernovae (SNe Ia) in the redshift range $0.01<z<2.26$ \cite{Brout:2022vxf}.}
\item{``Union3'' - comprising a \wlm{quadratic spline interpolation of 2087 SNe Ia, around a flat $\Lambda$CDM model, at 22 nodes} over the redshift range $0.05<z<2.26$, standardised and compiled using the Unity 1.5 Bayesian framework \cite{Rubin:2023ovl}.}
\item{``DES-SN5YR'' - from the Dark Energy Survey's 5-year result, including 194 historical SNe Ia in the low redshift range $0.025<z<0.1$ and 1635 SNe~Ia with photometric redshifts in the range $0.1<z<1.3$ \cite{DES:2024tys}.}
\end{itemize}
\wlm{The particular data we use are based on the publicly-available likelihood codes, implemented in \textsc{cobaya} and include magnitudes at various redshifts, along with covariance matrices.} These 3 datasets provide the very necessary empirical observations with which we may do cosmology. \wlm{Once again, we draw attention to the fact that Type Ia Supernovae undergo standardisation and fitting, according to a model, so they are only empirical in so far as this process is understood to have been done.}
The fact that each dataset is a distinct compilation and analysis of supernova data (even though the supernovae of the datasets are not necessarily completely distinct\footnote{Over 85\% (1363) of the supernovae from Pantheon+ are included in the Union3 compilation, and the 194 low-redshift supernovae from DES-SN5YR are also present in the other two \cite{desicollaboration2024desi}.}), while supposed to be produced by one underlying model (whether $\Lambda$CDM or otherwise), is ideal leverage to test their consistency with respect to each other.
\section{Template Functions}
\label{temps}
In this section we present the functions that will be used as template functions in the consistency analysis.
\subsection{Model-based template function}
\begin{figure}[h!]
\centering
\includegraphics[scale=0.42]{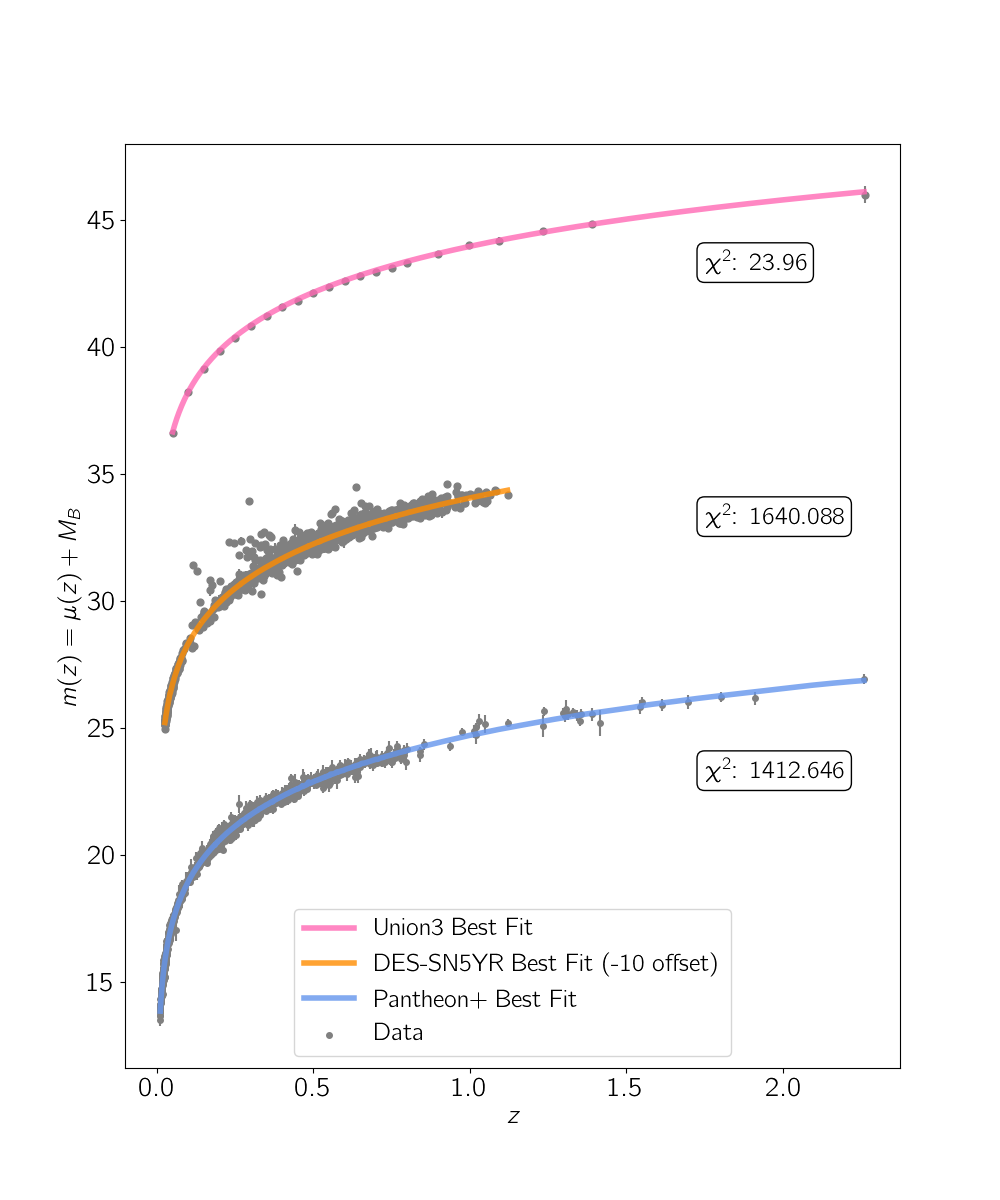}
\caption{Data and flat $\Lambda$CDM best fit curves of $m(z)$ for each of the data compilations considered. The components relevant to DES-SN5YR have been offset by an additive factor $10$ to facilitate viewing on the same axes. The offset between the other two datasets is due to the empirical difference in the individual values of $M_B$ of each.}
\label{bfmu}
\end{figure}

To begin, we perform a fit to the concordance flat $\Lambda$CDM model, using \textsc{cobaya}~\cite{Torrado:2020dgo} and MCMC sampling. Where necessary, we use the inferred parameter values from Planck 2018 Results~\cite{Planck1:2020}.
The best fit parameter values for each dataset are reported in Table \ref{bfres}, along with the corresponding $\chi^2$. The equivalent best fit distance moduli are shown in Fig. \ref{bfmu}, where we introduce the colour convention adopted in this paper. Anywhere where the result relates to the covariance and $\chi^2$ to data from Pantheon+, Union3 and DES-SN5YR, we will use the colours blue, pink and orange, respectively.
\begin{table}[h!]
\centering
\begin{tabular}{c|c c c}
Parameter & Pantheon+ & Union3  & DES-SN5YR\\
\hline
$\Omega_m$ &$0.33141\pm0.05384$&$0.35576\pm0.036427$&$0.35175\pm0.12699$\\
$M_B$ $(\Delta M_B)$ &$-19.420\pm0.155$&$-0.14894\pm0.09129$&$-0.057596\pm0.21322$ \\
\hline
$\chi^2$ &1412.6&23.960&1640.1\\
\wlm{$\chi^2/DoF$} &0.912&1.1980&0.946\\
\end{tabular}
\caption{Best fit results for the flat $\Lambda$CDM model parameters from each of the three datasets considered. Note that the much lower value of $\chi^2$ for Union3 is due to the fact that the data are \wlm{interpolated $2^{\rm nd}$ order splines around a flat $\Lambda$CDM model, at 22 node points}. In the case of Union3 and DES-SN5YR the data are also given with the $M_B$ value already taken care of, and so the small shifts found while fitting (referred to here as $\Delta M_B$) are very close to 0. \wlm{Here ``DoF'' refers to the number of degrees of freedom, which is the (effective) number of data points in each dataset minus the number of parameters fit for a particular model \cite{Brout:2022vxf,Rubin:2023ovl,DES:2024tys}.}}
\label{bfres}
\end{table}

As can be seen from the best fit parameter values in Table \ref{bfres}, there is no significant tension between the datasets; the inferred $\Omega_m$ values agree to within $1\sigma$ in all cases. As already mentioned, this does not preclude the existence of inconsistencies between the datasets, but is a necessary condition under the assumption that the model of the Universe is known, and that observations are actual realisations of it. 
These fits will be used as the $\Lambda$CDM model-based template functions in the subsequent consistency analysis. 
The applications of crossing functions to this template function allows us to test how much freedom is required for the best fit from one dataset to be a good fit to the other two, and whether there is any region in the space of hyperparameters that allows for the realization of all 3 datasets. This will allow us to gauge the level of consistency of the datasets, under flat $\Lambda$CDM and test the first null hypothesis.


\subsection{Model-independent template function}
\label{miTF}
\begin{figure}[H]
\centering
\includegraphics[scale=0.4]{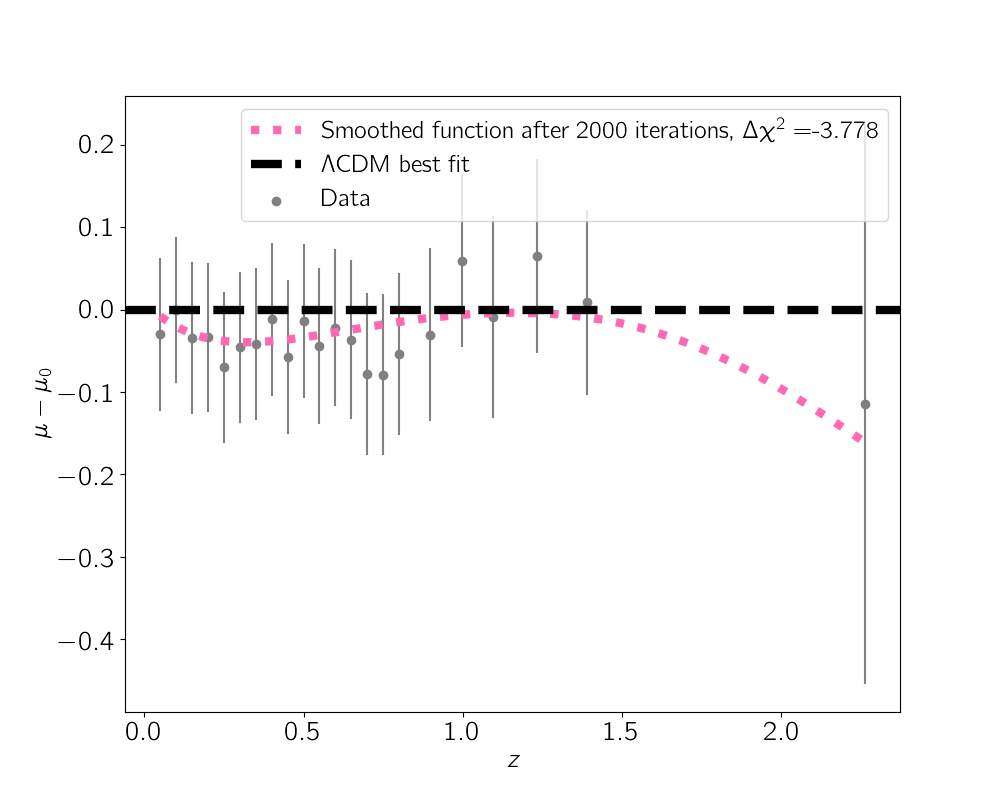}
\caption{Residuals of the iteratively-smoothed function for $\mu(z)$ with respect to the best fit from flat $\Lambda$CDM of Union3 data.}
\label{IS_u3}
\end{figure}

\begin{figure}[H]
\centering
\includegraphics[scale=0.355]{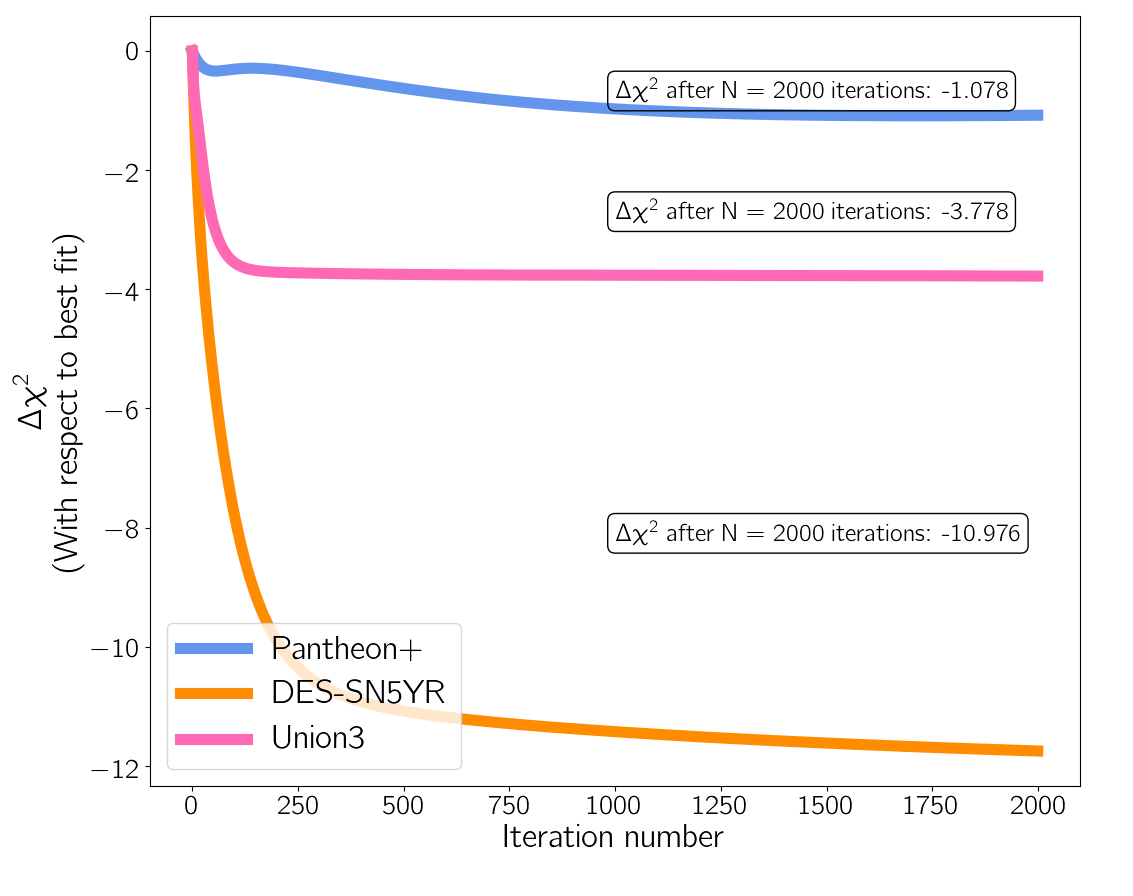}
\caption{Evolution of the $\Delta\chi^2$ with iteration number, taken with respect to the $\chi^2$ of the flat $\Lambda$CDM best fit of each dataset. We see that, although the improvement in $\chi^2$ for Pantheon+ is negligible (considering the large number of supernovae, and thus degrees of freedom), the improvement is about an order of magnitude more significant in the case of DES-SN5YR. The iterative smoothing function successfully converges on a closer fit to the data, and thus represents a reasonable model-independent function through the data.}
\label{IS_dchi}
\end{figure}

We employ iterative smoothing to generate a template function for $\mu(z)$ that is independent of the background cosmological model, \wlm{taking as our $0^{th}$ iteration, the $\Lambda$CDM best fit from each dataset}. In Fig. \ref{IS_u3} we show an example of this in the case of Union3 data, by plotting the residuals of the function resulting from iterative smoothing with respect to the cosmological model best fit\footnote{The keen-eyed reader might notice the asymmetric distribution of the residuals around the zero line corresponding to the flat $\Lambda$CDM best fit. This is not altogether unexpected, given that the \wlm{splined} Union3 data show strong off-diagonal correlations in the covariance matrix.}. As may be seen in the case of Union3, and is indeed true of all cases, the iterative smoothing procedure results in an overall improvement in the $\chi^2$ compared to the best fit function (see Fig. \ref{IS_dchi}), since it serves to create a smooth function that passes as close to as many data points as possible. For a fixed smoothing length $\Delta$, this will not necessarily converge on a function that passes through all the data points, since \wlm{there are only two parameters ($\Delta$ and $N$). Even with sufficiently many iterations,} $N\rightarrow\infty$, each part of the fitting function is controlled by some range of data points around it. Nevertheless, depending on the distribution of the $\mu$ data in redshift, the $\Delta$ and $N$ must be chosen in such a way as to avoid fitting small scale noise without being completely insensitive to the particular features of the dataset.


The $\Delta\chi^2$ from the improvement due to iterative smoothing is shown in Fig. \ref{IS_dchi} for all three datasets, with respect to their own, individual best fit $\chi^2$. Since there are only 22 data points for Union3, their density in redshift is lower than the other datasets, and so we adopt a slightly larger smoothing length. We also expect that, with fewer degrees of freedom, the change in its $\chi^2$ will be smaller in value than for the other two datasets. While this is indeed the case for DES-SN5YR, the improvement in Pantheon+ is the smallest overall. This is most likely due to the fact that the data are already very-well constrained by the best fit flat $\Lambda$CDM model, and consequently an iteratively-smoothed function cannot be found that drastically improves it, to the same level as in the case of the other datasets. 
See Section \ref{tuning} for a discussion of the choices of hyperparameters used in the generation of these template functions.

The resulting iteratively-smoothed functions will be used as template functions in the second part of the Crossing Statistics analysis. Since each one is constructed \wlm{to more closely} fit to its own dataset, they will have taken up some characteristics of their dataset, in a model-independent way. 
The crossing functions which are then applied after smoothing will allow for large-scale fluctuations of each template function that will slightly deform (and in a sense relax) the characteristics from its own dataset, and determine to what extent these peculiarities can remain, or not, in order to be a good fit to the other datasets. This allows us to test the second null hypothesis.

\vspace{-5pt}

\section{Implementing Crossing Functions}
\label{crossstatmeth}
\subsection{$\Lambda$CDM best fit}

Here we show the results of the application of Crossing Statistics, using the flat $\Lambda$CDM best fits to each of the three datasets as the template functions. \wlm{The posteriors of the deformed functions of $\mu(z)$ (generated by MCMC sampling in the space of crossing hyperparameters, and $M_B$)} are then calculated for each of the three datasets, to determine the possible shapes of $\mu(z)$ suggested by the data. Using the $\chi^2$ calculated in this manner, we may determine how the deformations at various orders might improve the fit to the data, and the effect of using the best fit from a different dataset.

\begin{figure}[h!]
\centering
\includegraphics[scale=0.4]{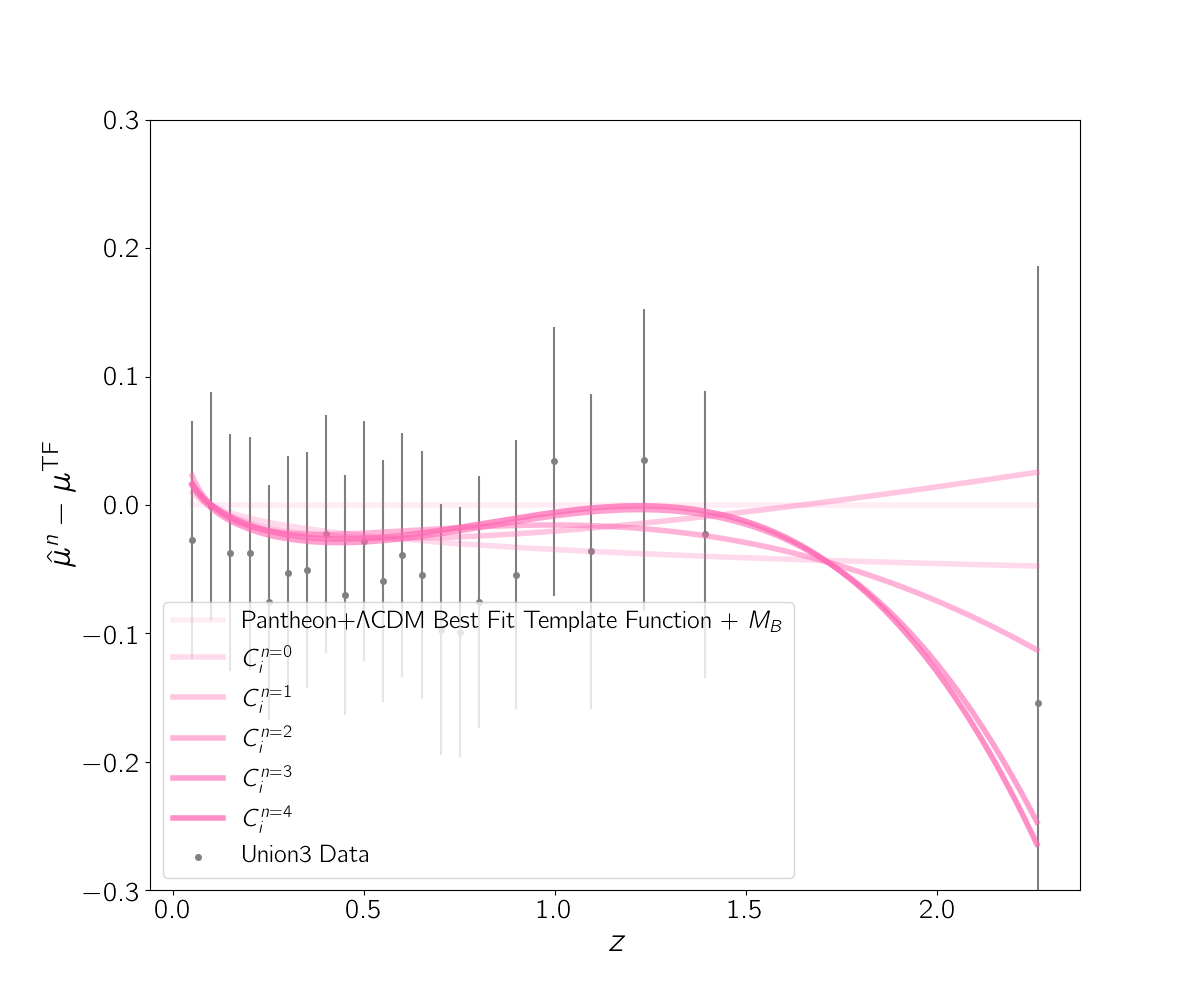}
\caption{Residuals of the deformations to the template function originating from the Flat $\Lambda$CDM fit to Pantheon+ data, with respect to the original (undeformed) version. The deformations (shown in pink) include crossing hyperparameters, up to various maximum orders $n$ shown in the legend, whose values are chosen to give the best fit to the Union3 data (shown in grey).}
\label{n0recon}
\end{figure}

\wlm{In Fig. \ref{n0recon}, `$C^{n=k}_i$' refers to the case where the template function for distance modulus is deformed by the Chebyshev polynomials up to order $k$}. In the case where we only have the template function no crossing functions are used, but the absolute magnitude $M_B$ is varied as a nuisance parameter.\footnote{Note: Absolute magnitude is not completely degenerate with $C_0$. $M_B$ is a translation (additive) in $\mu$, while $C_0$ is a multiplicative factor.} As an example, we show how the application of crossing functions deforms the Pantheon+ best fit template function to better fit the Union3 data. 

At each increasing order in Crossing Statistics, the deformed template function is allowed more freedom, and so there is a necessary improvement in the $\chi^2$ compared to lower orders (and the template function). For a summary of the initial and best $\chi^2$ values, see Table~\ref{BFchi2}. In this table, we also introduce for the first time our abbreviations: for the datasets, to which the deformed functions are fit, we use ``P+'', ``U3'', ``D5'' for Pantheon+, Union3 and DES-SN5YR respectively, and we abbreviate 'template function' to ``T.F.''.

\begin{figure}[h!]
\hspace{-65pt}\includegraphics[scale=0.33]{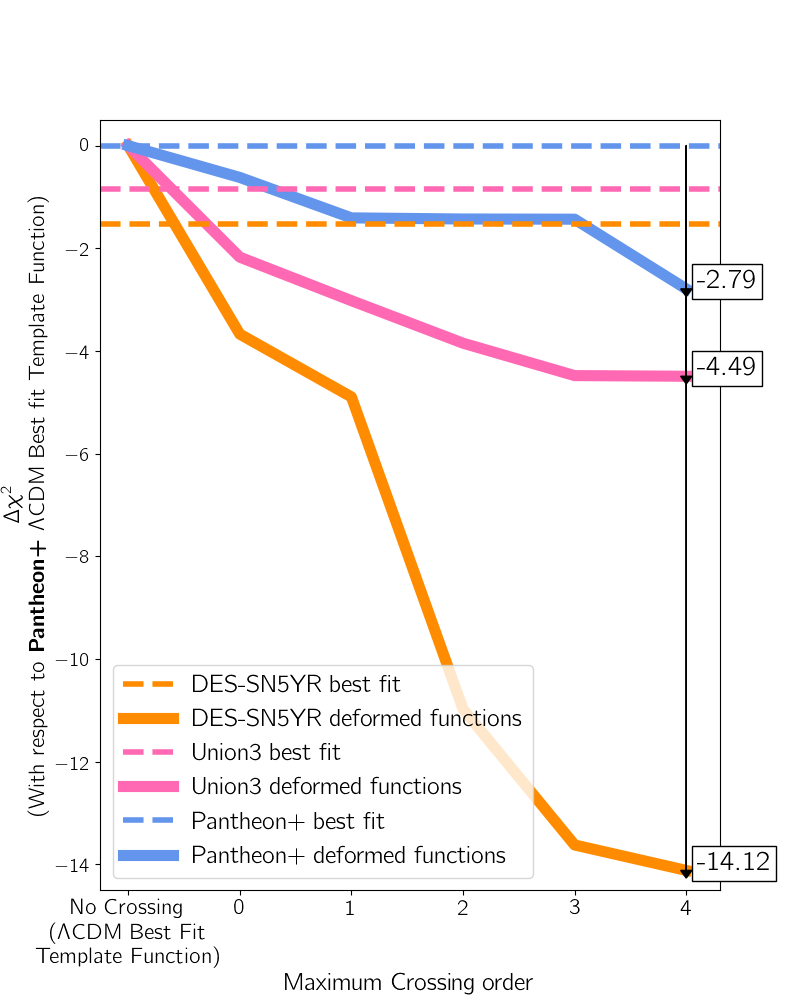}\includegraphics[scale=0.33]{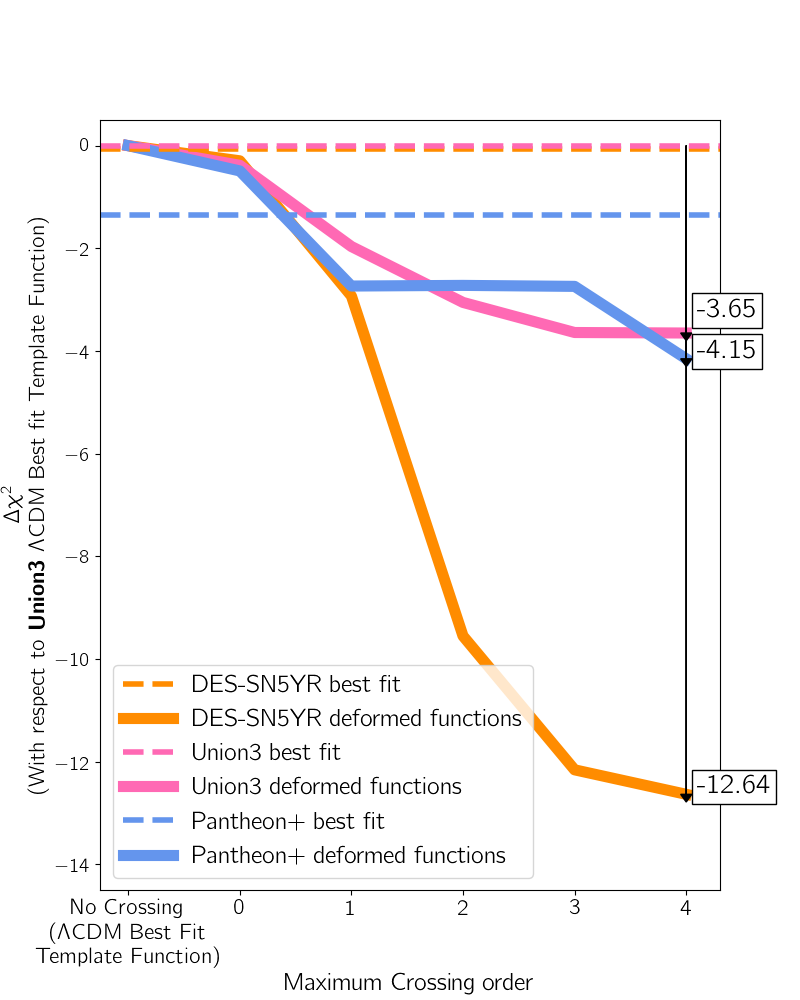}\includegraphics[scale=0.33]{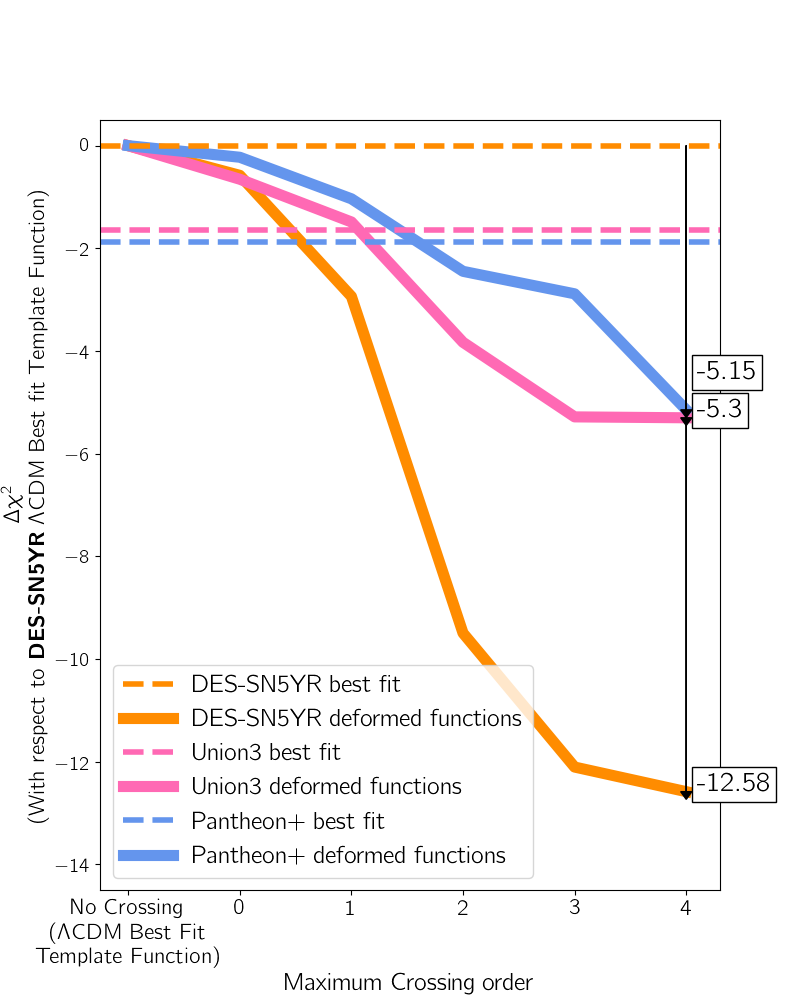}
\caption{\textit{Left panel}: The $\Delta\chi^2$ improvement for each deformation is shown with a solid line, and the dashed lines correspond to the $\Delta\chi^2$ obtained using the best fit flat $\Lambda$CDM template function from each dataset, according to our colour convention. All $\Delta\chi^2$ are taken with respect to the template function (Pantheon+ Flat $\Lambda$CDM best fit) $\chi^2$ value. We also report the $\Delta\chi^2$ achieved from the $4^{th}$ order crossing functions. \textit{Middle panel}: The  analogous figure, but using the Union3 flat $\Lambda$CDM best fit as the template function. Note: the $\Delta\chi^2$ of the DES-SN5YR best fit is quite similar to that of Union3, and is difficult to see on the plot. \textit{Right panel}: The analogous figure, but where DES-SN5YR flat $\Lambda$CDM best fit is used as a template function.} 
\label{n0dchi}
\end{figure}

In Fig. \ref{n0dchi}, we see how the $\Delta\chi^2$ evolves with maximum order of the applied crossing functions\wlm{, from ``No Crossing'' (undeformed template function) to $4^{\rm th}$ order (with 5 additional free parameters)}. \wlm{This is calculated as $\Delta\chi^2 = \chi^2\left[\hat \mu^{TF, A}(z) \sum_{i=0}^{n} C^n_i T_i(\tilde z)\right]_B - \chi^2\left[\hat \mu^{TF, A}(z)\right]_B$, i.e. the difference in $\chi^2$ between a deformed template function A, fit to dataset B, and the undeformed version of the same template function A, fit to the same dataset. In this sense, it shows the decrease in $\chi^2$ as more freedom\footnote{For a discussion concerning the degrees of freedom and over-fitting, see Appendix~\ref{tuning}} is given to the deformation, by increasing the number of hyperparameters. For each panel, the undeformed template function A is fixed, and stated on the y-axis. The curves in a particular panel then correspond to the deformations (at various orders, stated on the x-axis) of this template function, to each each dataset. In the ``No Crossing'' case corresponds to the undeformed template function, it is always identically zero and is shown for reference.}

It is also important to consider the relative $\Delta\chi^2$ that may be achieved by simply using the template function of the data itself (without application of Crossing Statistics), \wlm{i.e. $\Delta\chi^2 = \chi^2\left[\hat \mu^{TF, B}(z)\right]_B - \chi^2\left[\hat \mu^{TF, A}(z)\right]_B$.} These are shown as a dashed horizontal lines, coloured according to the corresponding dataset used to generate the template function. 

The results (\wlm{in the \textit{left} and \textit{middle} panels}) from Union3 and Pantheon+ template functions ($\mu^{\rm TF}$) show that, even at the $0^{th}$ order in Crossing Statistics (i.e. a deformation according to $(\mu^{TF}(z)+M_B)\times C_0$), the deformed function can outperform the Flat $\Lambda$CDM best fit from the data, even in the case where the best fit template function does not come from the data itself. This would seem to indicate that the differences between the datasets are not so significant as to prevent the freedoms in one dataset from being explored by a best fit constrained on another dataset. This independence of the new $\chi^2$ from the deformed function on the dataset used to determine the best fit points towards a good acceptance of these datasets to the others' flat $\Lambda$CDM best fit. 

This is true for all these cases, except where the Union3 template function is used in conjunction with the Pantheon+ data \wlm{(\textit{middle} panel, blue curve)}. Here, it is only at $1^{st}$ order in Crossing Statistics that the improvement from using the deformed Union3 template function, surpasses that of the best fit from Pantheon+ (with respect to the undeformed Union3 template function). This means that the Union3 template function is \textit{not} intrinsically as good a fit to the Pantheon+ data, if given only the 2-parameter freedom of: $M_B$ and $C_0$, whereas the Pantheon+ template function \textit{is}, when fit  the Union3 data\footnote{This may simply be due to the fact that the Pantheon+ dataset has many more points in redshift, and so more freedom is required by the Union3 template function to surpass the Pantheon+ best fit.}.

By contrast, if we consider the DES-SN5YR best fit template function (\wlm{\textit{right} panel}), we see that it requires the flexibility provided by at least 2$^{nd}$ order in crossing functions before it is able to \wlm{obtain a lower $\chi^2$ in the context of} the \textit{other} two datasets \wlm{(pink and blue curves)}. It is difficult to draw conclusions from the $\chi^2$ alone, but this hints that the DES-SN5YR best fit may be slightly less compatible with the other two datasets, for some reason\footnote{As in the case of the Union3 template function, this may be a result of differences in the characteristics of the data, such as the redshift range or sparseness of the data points in redshift, but conclusions cannot be drawn from the $\chi^2$ value alone.}.

\begin{table}[H]
\small
\centering
\begin{tabular}{c|c c c|c c c|c c c}
Flat $\Lambda$CDM T.F.: &\multicolumn{3}{c|}{Pantheon+}&\multicolumn{3}{c|}{Union3}& \multicolumn{3}{c}{DES-SN5YR} \\
Data: & P+ & U3  & D5& P+ & U3  & D5& P+ & U3  & D5\\
\hline
T.F. (with $M_B$)&1412.65&24.8091&1641.61&1414.00&23.9600&1640.16&1414.51&25.6061&1640.09\\
$4^{th}$ order& 1409.86&20.3172&1627.49&1409.72&20.3414&1627.70&1409.37&20.3052&1627.51
\end{tabular}
\caption{$\chi^2$ achieved from $4^{th}$ order crossing functions applied to the template functions (T.F.'s) from the Flat $\Lambda$CDM best fit, compared to fitting just a translational factor $M_B$. We see here that the maximum $\Delta\chi^2$, achieved by application of crossing functions at $4^{th}$ order, is practically independent of the chosen template function.}
\label{BFchi2}
\end{table}

\subsection{Iterative Smoothing}
\label{itsmuthmeth}
We perform a similar analysis to the previous section, but where the template functions are now the iteratively-smoothed functions of each dataset. We use the iterative smoothing functions only from the Union3 and Pantheon+ datasets, as template functions for each of the three datasets\footnote{Here we do not use DES-SN5YR to make a template function for the other datasets, as the data are only present until $z\sim1$, and since we are not using a model, the template function cannot be trivially extended to the higher redshifts necessary to sample the $\mu$ corresponding to the other datasets.}. Since the template function comes from a different dataset, we include the best fit value of $M_B$ (i.e. translational freedom in the y-axis direction) so that the template function can shift to lie in the appropriate range. 

Since each template function is now one which is iteratively-smoothed to more closely conform to its particular dataset, it will not necessarily immediately be a particularly good fit to any other dataset. Thus, when the crossing functions are implemented on it, to fit another dataset, we cannot expect that the $\chi^2$ will necessarily drop below the $\chi^2$ using the latter dataset's own iteratively-smoothed function. As a result, we see that in the cases where one dataset's iteratively-smoothed function is deformed with crossing functions chosen to fit another dataset, a larger maximum order (of at least 3) is required to recover as good a fit as the iteratively-smoothed function of the other dataset itself. The initial and best $\chi^2$ results from this part of the analysis are summarised in Table~\ref{ISchi2}. To get the full picture about mutual consistency, we must examine the posteriors of the corresponding hyperparameter values.
\begin{table}[H]
\small
\centering
\begin{tabular}{c|c c c|c c c|c c c}
Iterative Smoothing T.F.: &\multicolumn{3}{c|}{Pantheon+}&\multicolumn{3}{c|}{Union3}& \multicolumn{3}{c}{DES-SN5YR} \\
Data: & P+ & U3  & D5& P+ & U3  & D5& P+ & U3  & D5\\
\hline
T.F. (with $M_B$)&1411.57&25.0631&1644.79&1425.60&20.1819&1637.66&-&-&1629.03\\
$4^{th}$ order&1409.80&20.1784&1627.52&1411.02&19.9952&1627.53&-&-& 1627.52
\end{tabular}
\caption{$\chi^2$ achieved from $4^{th}$ order crossing functions applied to the template functions from iterative smoothing, compared to using just a translational factor $M_B$. Once again we see here that the maximum $\Delta\chi^2$, achieved by application of crossing functions at $4^{th}$ order, is practically independent of the chosen template function.}
\label{ISchi2}
\end{table}

\wlm{It may be useful here to recall the number of data points and free parameters in each crossing model. The fit, or generated, template function is fixed in each test. The number of data points of interest are from the dataset to which the template function is deformed:
\begin{itemize}
\item Pantheon+: 1550 \cite{Brout:2022vxf}
\item Union3: 22 \cite{Rubin:2023ovl}
\item DES-SN5YR: 1829 (effective number: 1735, due to treatment of contamination) \cite{DES:2024tys}.
\end{itemize}
Each order in crossing has a number of free parameters equal to the maximum order, plus the free parameters: $M_B$ (translational freedom) and $C_0$ (multiplicative bias).
Hence, the number of free parameters in fitting a fixed template function to a particular dataset at $n^{\rm th}$ order are: $n+2$, i.e. in the range $2-6$ for the orders considered in our analysis.}
\section{Results and Discussion}
\label{DnR}
\subsection{Full redshift range}

\wlm{In this section, we report the initial results found when considering all three datasets across the full range of each in redshift. We note that the redshift range of the DES-SN5YR dataset is approximately half that of the other two, and thus requires some additional consideration. For each additional crossing hyperparameter, we use a Gaussian prior over the range encompassing their standard values, with 10 on either side, e.g. $C_0\in[-10,10]$. This is done to prevent large deviations or unphysical behaviours coming from prior volume effects, assuring that those deviations we see are data-driven. Though this prior might superficially seem the same for each combination of template function and data, the fact that the DES-SN5YR redshift range is smaller means that the effective prior, or the interpretation of the hyperparameters, could be different. The most obvious reason is that the $z_{max}$ used in the Chebyshev polynomials is the overall maximum of all the datasets. This means that deforming the DES-SN5YR template function, in the interval $\tilde{z}\in[-1,1]$ of the Chebyshev polynomials, affect the template function at redshifts not directly informed by the original data.}

\wlm{Vice versa, when fitting the other template functions to the DES-SN5YR dataset, it only has constraining power over a fraction of the range spanned by the Chebyshev polynomials, and so it is reasonable to expect that the allowed values of the hyperparameters will cover a larger range (wider contours) to allow stronger deformations, especially at higher redshifts.}

\wlm{The analysis in this Section is thus most relevant for the comparison of Union3 and Pantheon+ across the full range of available redshifts, though we include DES-SN5YR to have a baseline that will allow us to assess the effect of these differences through further analysis (see Section~\ref{lowzmax}).}

\subsubsection{$\Lambda$CDM best fit template function}

\begin{figure}[H]
\centering
\includegraphics[scale=0.33]{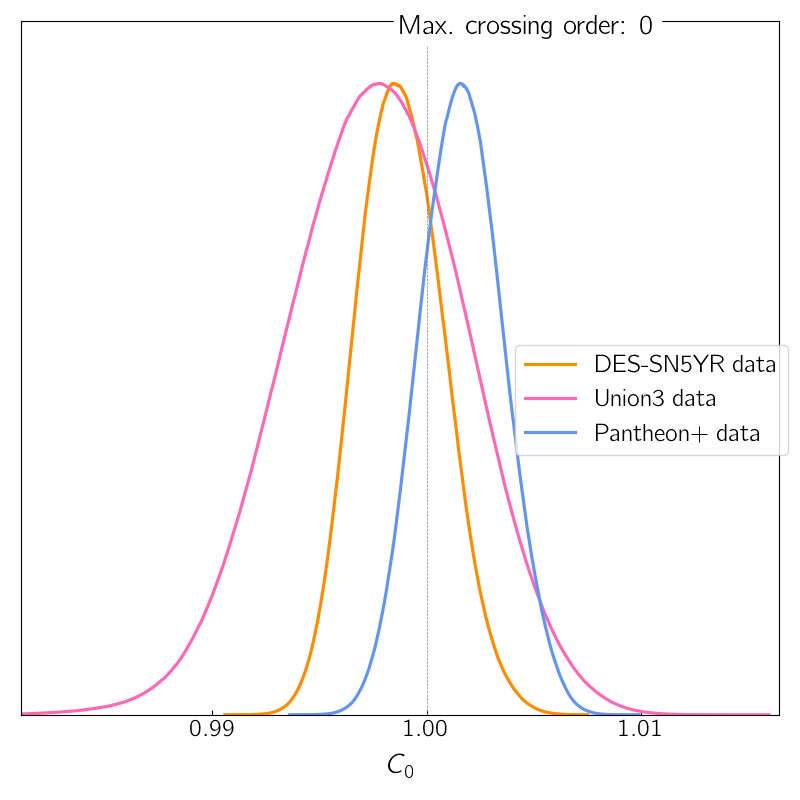}\includegraphics[scale=0.2]{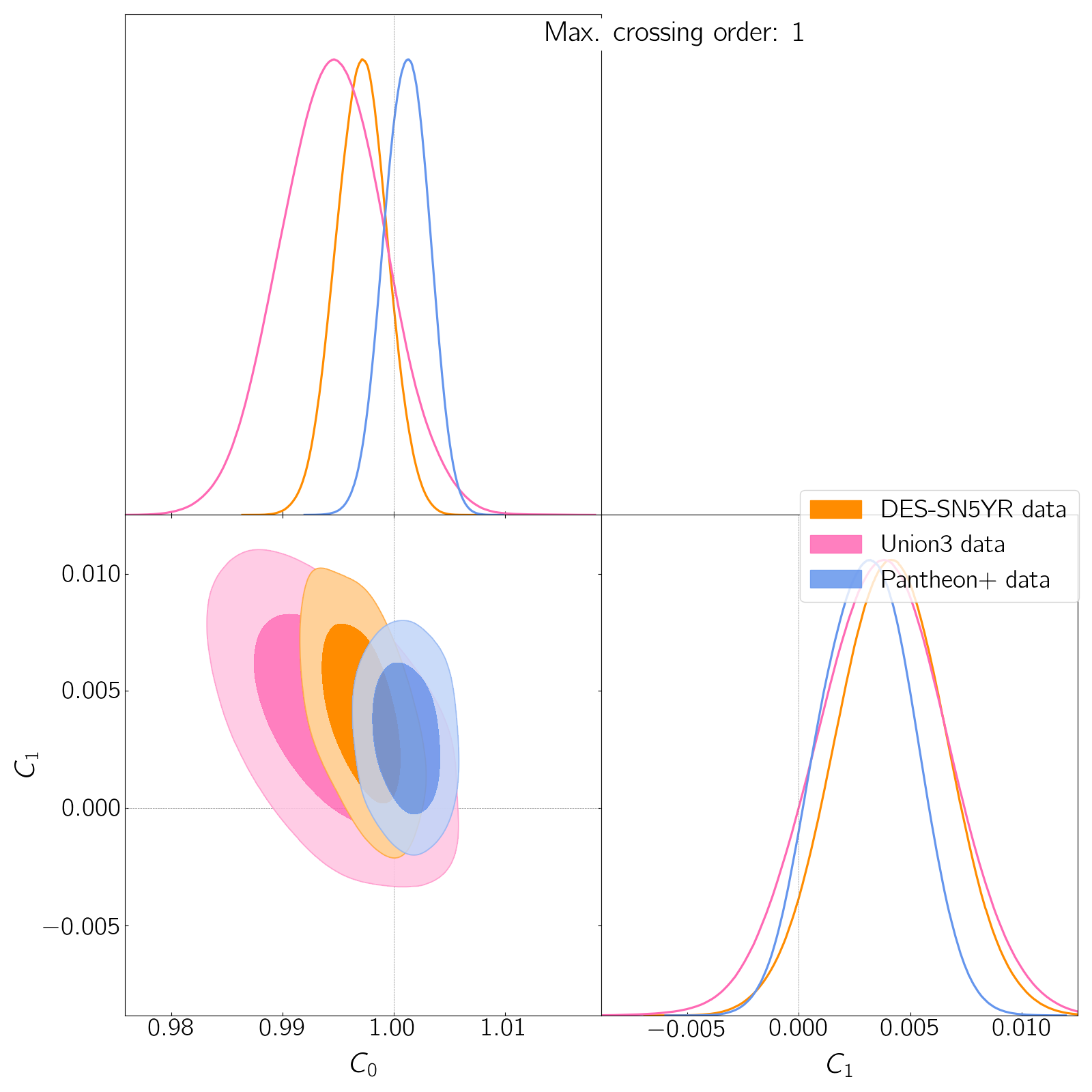}
\includegraphics[scale=0.19]{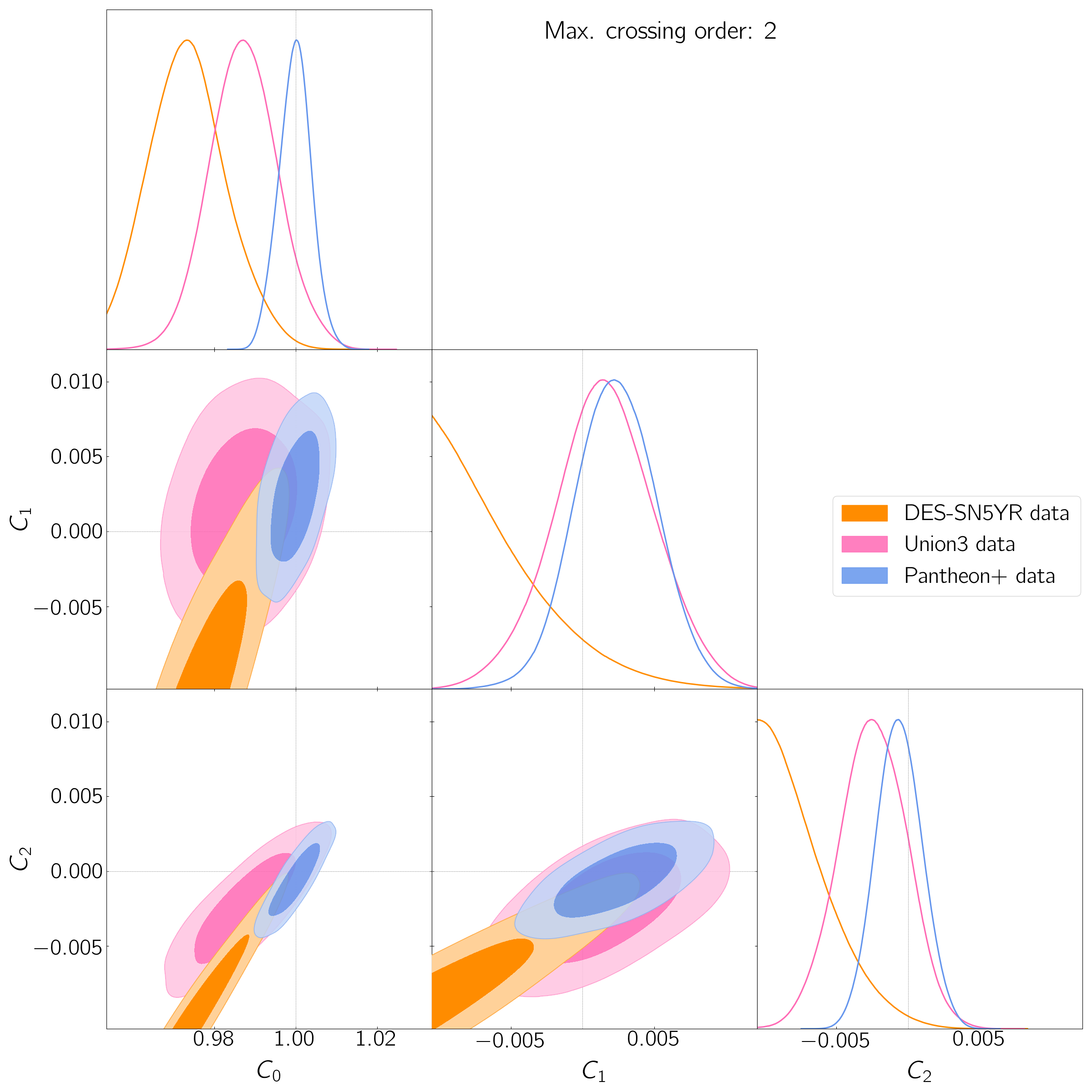}
\caption{Contour plots for the hyperparameters of the \textbf{Union3 $\Lambda$CDM best fit} template function, deformed to fit to all datasets. The contours from fits to Pantheon+, Union3 and DES-SN5YR data are shown in blue, pink and orange, respectively. Clockwise from the upper left panel, we show the results at maximum crossing function order of $0$, $1$ and $2$.}
\label{U3BFcont}
\end{figure}
\begin{figure}[H]
\centering
\includegraphics[scale=0.33]{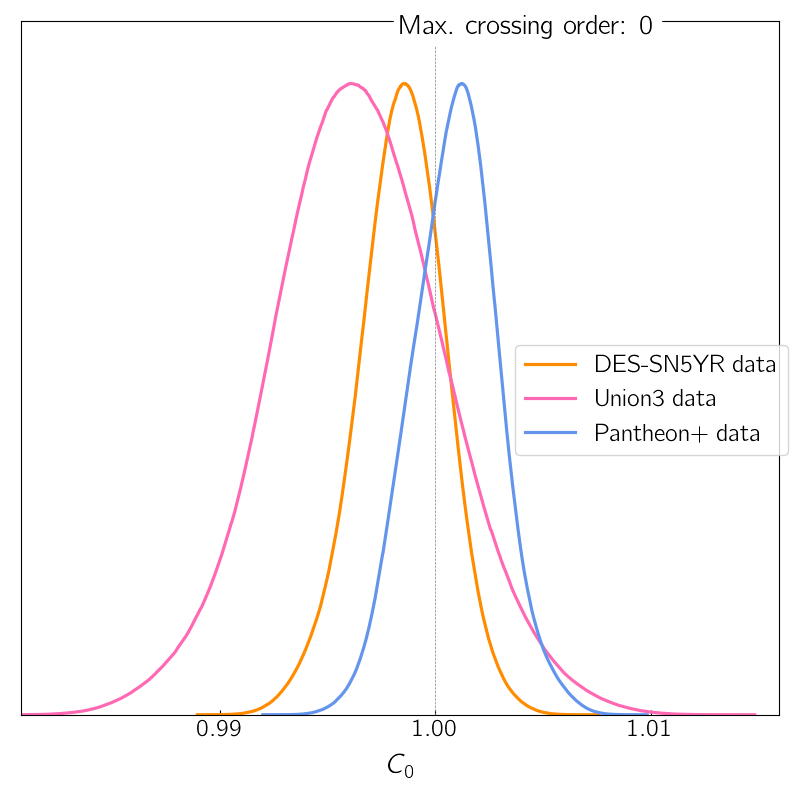}\includegraphics[scale=0.2]{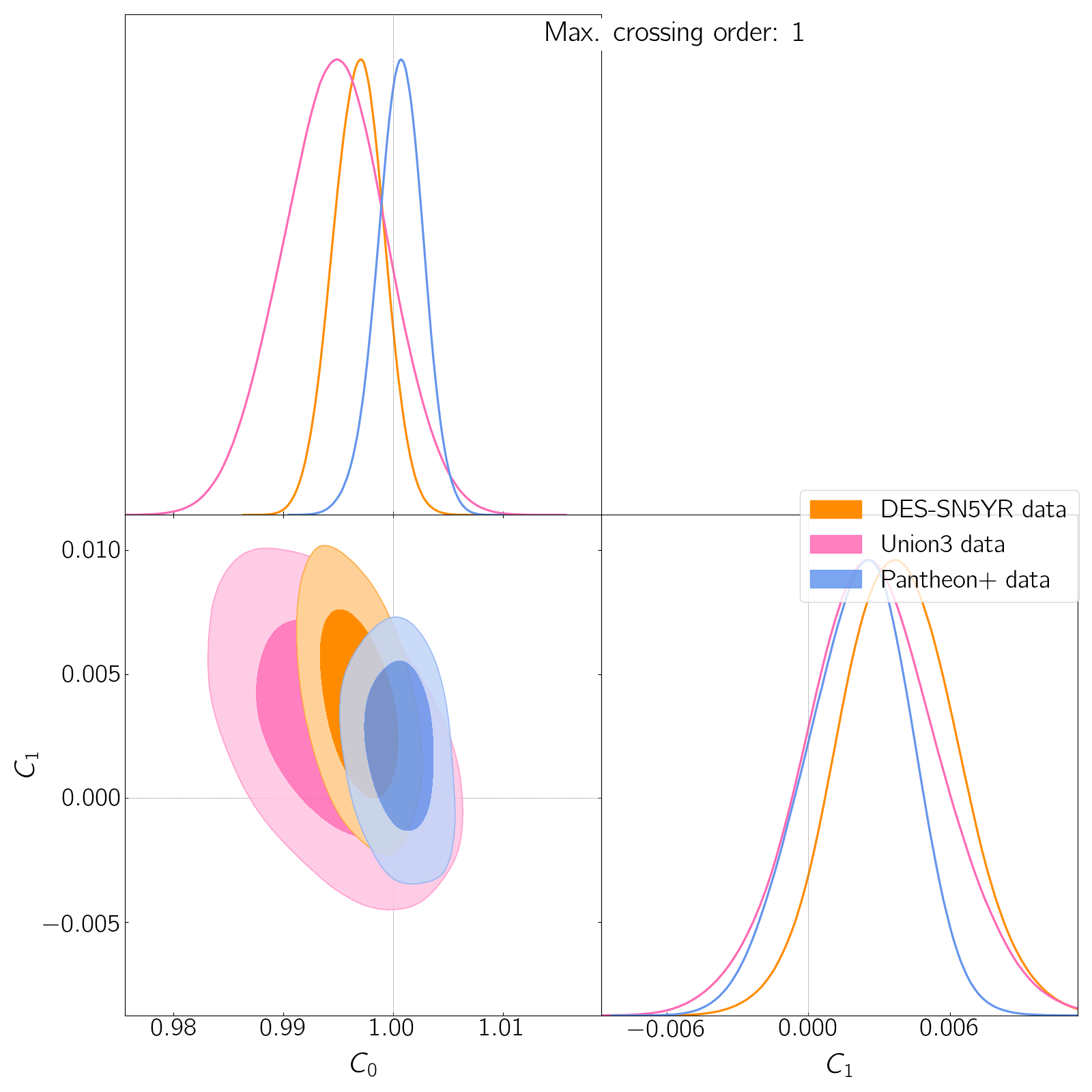}
\includegraphics[scale=0.19]{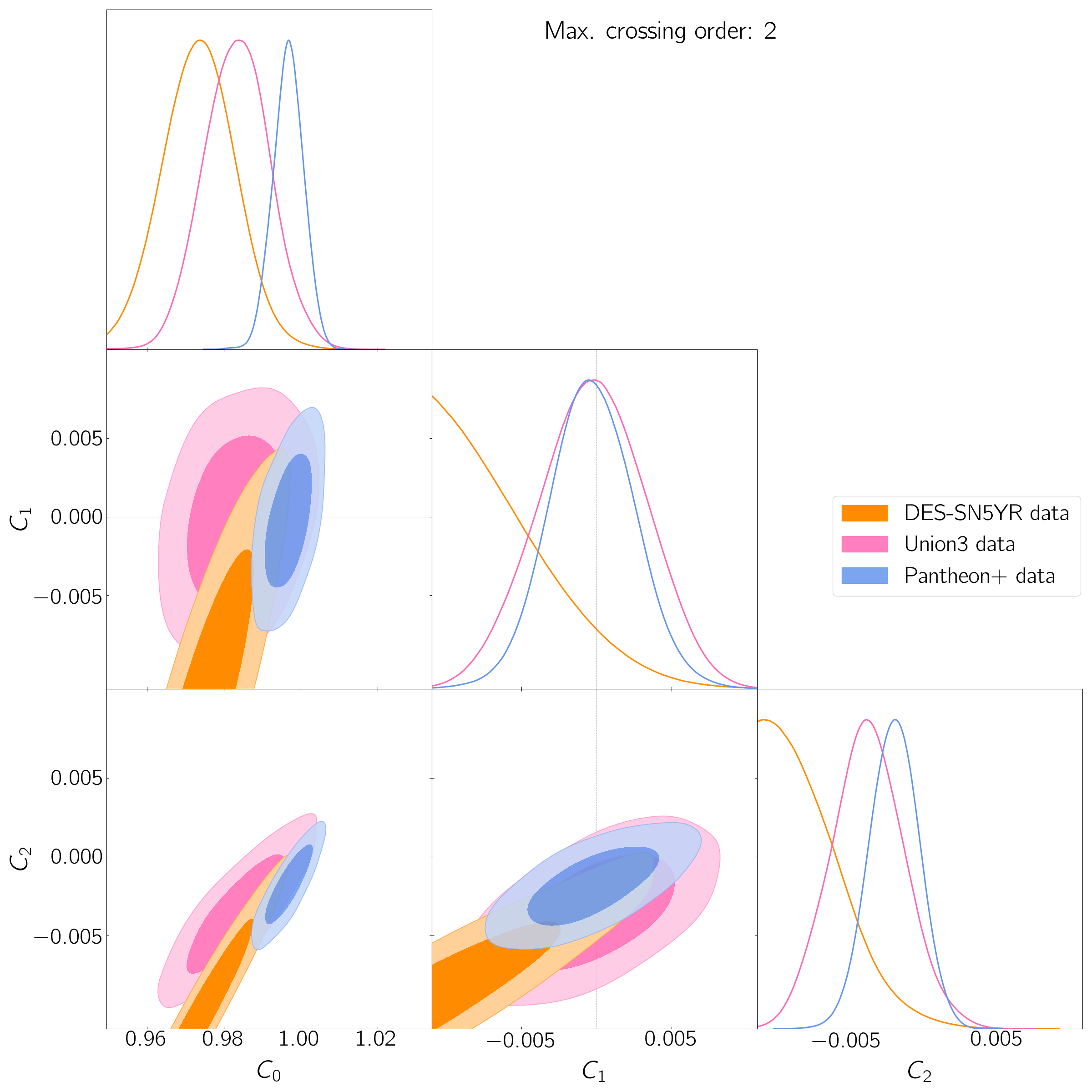}
\caption{Contour plots for the hyperparameters of the \textbf{DES-SN5YR $\Lambda$CDM best fit} template function, deformed to fit to all datasets. The contours from fits to Pantheon+, Union3 and DES-SN5YR data are shown in blue, pink and orange, respectively. Clockwise from the upper left panel, we show the results at maximum crossing function order of $0$, $1$ and $2$. We see that, even in this case where we make use of the DES-SN5YR best fit to $\Lambda$CDM as the template function, the hyperparameters still prefer values corresponding to deformations away from the concordance model, as opposed to the other two datasets.}
\label{D5BFcont}
\end{figure}

We show, in Figures~\ref{U3BFcont} and~\ref{D5BFcont}, some example contour plots for the hyperparameters used for the crossing functions. We focus on the cases of the Union3 and DES-SN5YR template function, as the two latest, distinct supernova compilations, but the overall shape and behaviour of the contours for Pantheon+ template functions are similar. As elsewhere, we refer to the values $(C^n_0,C^n_1,...,C^n_n) = (1,0,...,0)$ as the standard hyperparameter values at $n^{th}$ order in crossing functions, since they imply deformations to the template function effectively corresponding to identity. We show results up to $2^{nd}$ order in crossing functions, with the remaining relevant orders for the Union3 and DES-SN5YR template functions shown in Appendix~\ref{cont}.

In the case of the Union3 template function, for $C^{0}_{i}$ to $C^{2}_{i}$, there is about $1\sigma$ agreement for the hyperparameters of all datasets with the standard values (see Table~\ref{BFcontsum} for the summary of these results). Conversely, the Pantheon+ template function shows good \wlm{compatibility} with its own data, but for the other datasets the hyperparameters are only within $2\sigma$ of the standard values. At 3rd order in Crossing Statistics Union3 and Pantheon+ are both \wlm{compatible} both ways (as template function or data) up to $1\sigma$, but DES-SN5YR data is still more than $2\sigma$ away from the standard values. 

To understand this, we examined the contours where the DES-SN5YR best fit template function is used. At 0$^{th}$ and 1$^{st}$ order in Crossing Statistics the data do not prefer any behaviour away from the template function, and the hyperparameters are \wlm{compatible} with standard values to within $1\sigma$. However, for higher orders, as was seen for the template functions from Union3 and Pantheon+, the data only allow the standard hyperparameter values at $2\sigma$ or more. This is true even for the DES-SN5YR data itself, which means that it prefers behaviour that includes deformations away from its own best fit. In contrast, Pantheon+ data manages to agree with the DES-SN5YR best fit template function to within $1\sigma$ of the standard hyperparameter values. 

Once 4th order in Crossing Statistics is reached (and \wlm{the preferred deformations of the} DES-SN5YR \wlm{template function, when fit to DES-SN5YR data, now include the standard hyperparameters at less than $2\sigma$.}), all datasets are mutually \wlm{compatible} for the template functions tested. Note that the contours of the hyperparameters fit to DES-SN5YR data are significantly larger than for Union3 or Pantheon+ data, above $1^{st}$ order in crossing functions. 

It is useful to re-examine the $\Delta\chi^2$ in this context. In almost all cases from Union3 and Pantheon+ template functions (except the 1st order deformation of Union3, fit to Pantheon+
\vspace{-\textheight}\hspace{-25pt}data\footnote{Again, this implies Union3 $\Lambda$CDM best fit is slightly more difficult to match to the Pantheon+ data, i.e. requires more freedom.}), the $\Delta\chi^2$ already shows improvement over each dataset's own best fit at order corresponding to $C^{0}_i$. This means that they are not only consistent, but also fit the data well. 

When the DES-SN5YR template function is used, the $\chi^2$ to the other datasets only improves over the intrinsic best fit once the freedom from deforming with crossing functions of at least $2^{nd}$ order is allowed. It is also at this order that the Union3 and DES-SN5YR data prefer hyperparameter values that move the function away from the best fit. At lower orders, even though the undeformed best fit is the preferred behaviour, the fit is not as good. 

Thus, DES-SN5YR presents a slight difference to the other datasets. When its best fit is used as its own template function\footnote{This is true even when even when $z_{max}$ of $\sim1.1$ is used in the Chebyshev polynomials, see Section~\ref{lowzmax} for the detailed analysis of this case, and Appendix~\ref{cont}, Table~\ref{lowzBFcontsum} for the summary of results.}, \wlm{compatibility} with the standard hyperparameter values is only present at $0^{th}$ and $1^{st}$ order, then not re-attained until $4^{th}$ order. The cases of Union3 and Pantheon+ deformed template functions fit to the DES-SN5YR data prefer non-standard hyperparameter values at lower orders too. This indicates that, until $3^{rd}$ order, there is still some trend, other than $\Lambda$CDM, that is preferred by the data, (but in a way that is seemingly not \wlm{compatible} with the other datasets' best fit template functions, without deformation\footnote{See Section~\ref{IScontours} where we use iteratively-smoothed template functions to explore features of each dataset beyond the concordance model.}. At $4^{th}$ order, a lower $\chi^2$ may be obtained by values of hyperparameters \wlm{compatible} with standard values to within $1\sigma$, only because the uncertainties are large enough to include the standard hyperparameter values.

Beyond agreement with standard hyperparameter values, it is also important to assess whether there are regions of the hyperparameter space which all of the data prefer. In this case, there is always an overlap of around $1\sigma\sim2\sigma$, between the contours of all three datasets for all orders in crossing functions. This means that there are still (despite the increased sensitivity to other behaviours) expansion histories of some underlying universe that can produce all 3 datasets simultaneously, to within $1\sigma\sim2\sigma$.

\subsubsection{Iterative smoothing template function}
\label{IScontours}


In this part of the analysis we examine the effect on mutual consistency when the best fit template functions are replaced by iteratively-smoothed functions which \wlm{more closely follow the data from} each dataset and are model-independent. With the maximum redshift of some data exceeding that of DES-SN5YR, and no trivial way to extend an iterative smoothing function beyond the data range, only Union3 and Pantheon+ data are used to produce template functions with iterative smoothing\footnote{See Section~\ref{lowzmax} where we consider a different redshift range in all data.}.

\begin{figure}[H]
\centering
\includegraphics[scale=0.33]{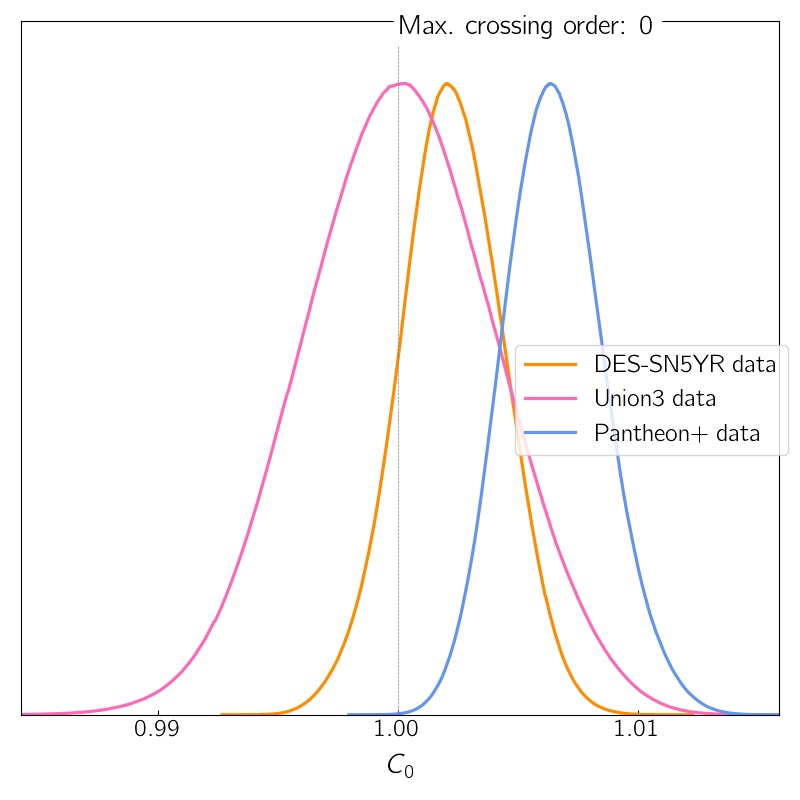}\includegraphics[scale=0.2]{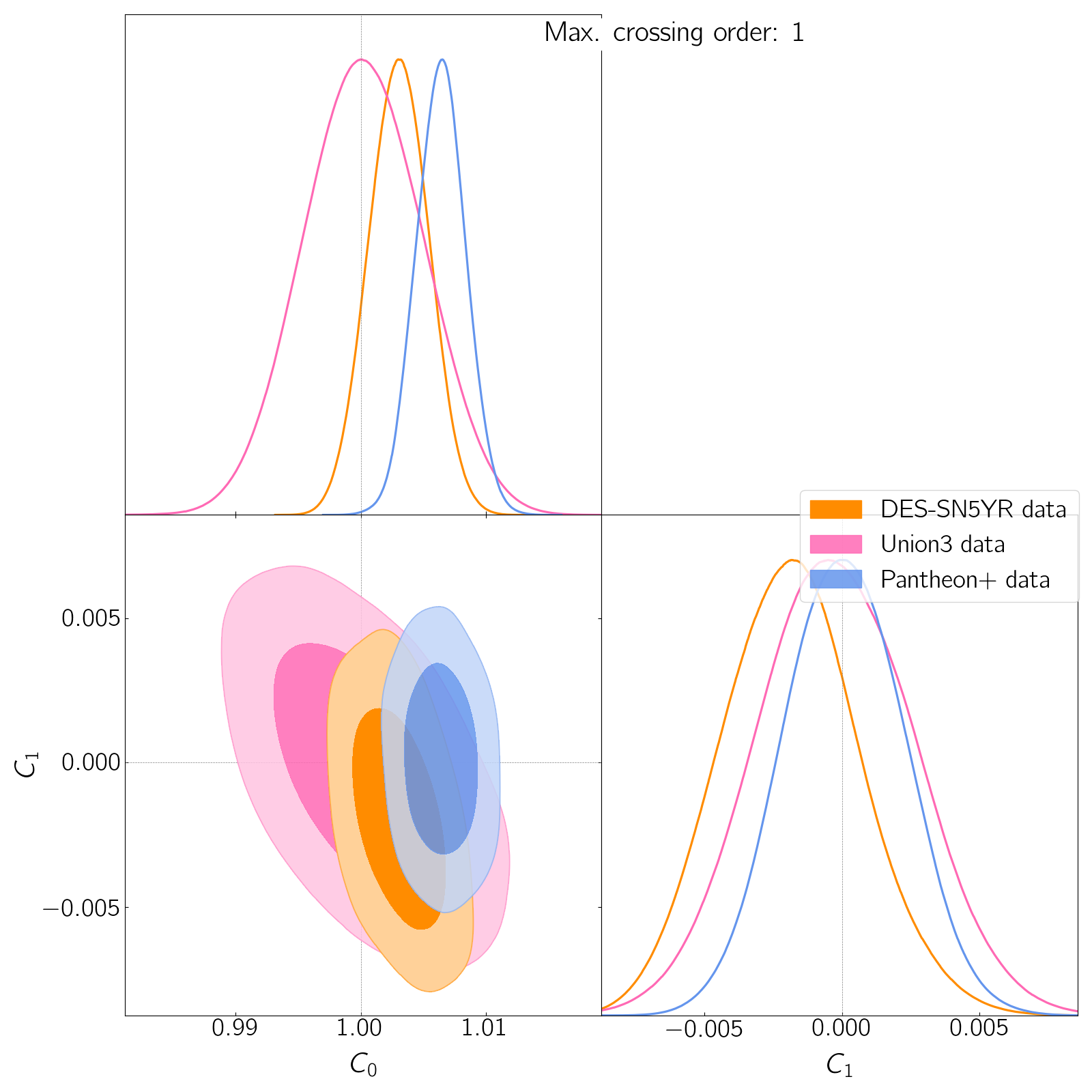}
\includegraphics[scale=0.19]{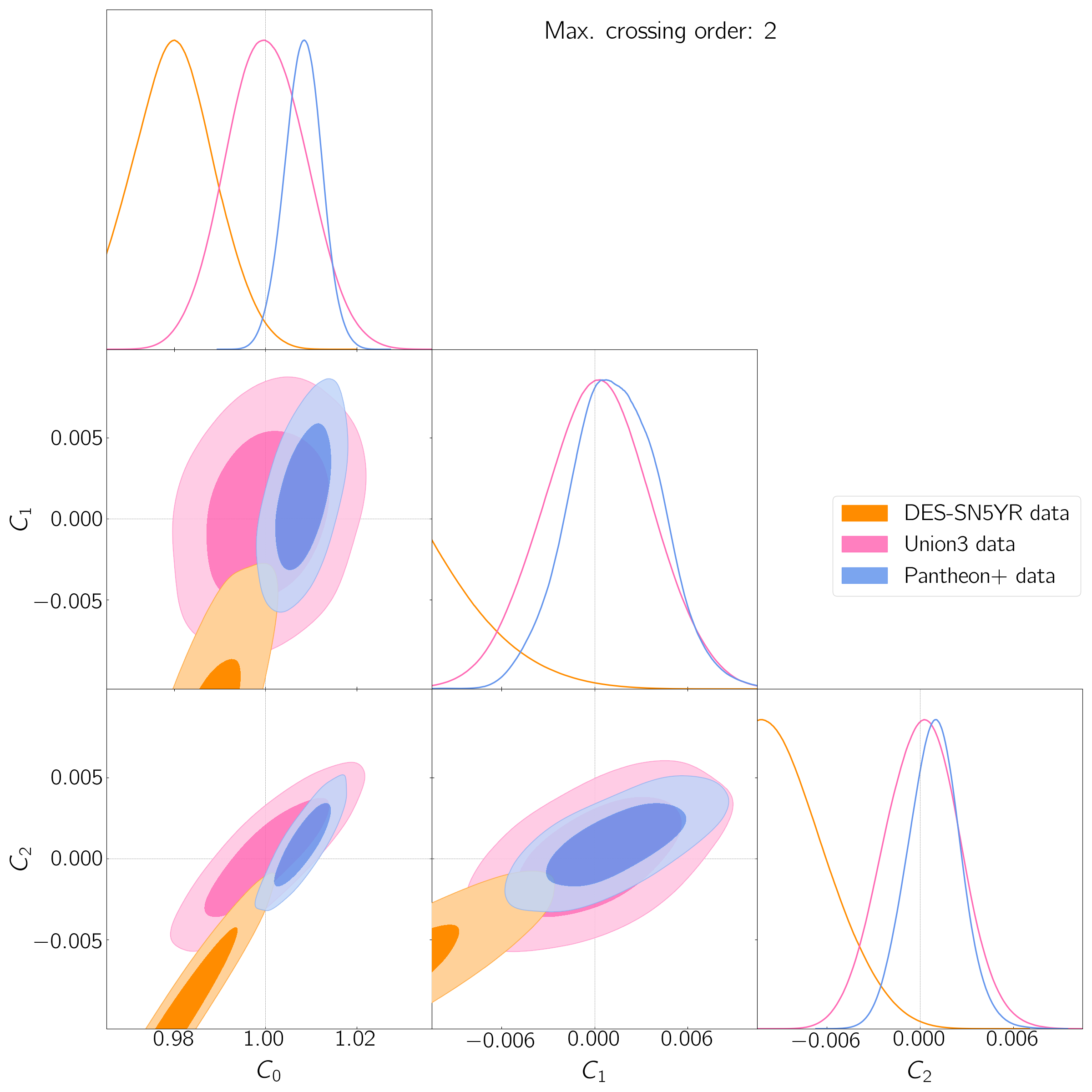}
\caption{Contour plots for the hyperparameters of the \textbf{Union3 iterative smoothing} template function, deformed to fit to all datasets. The contours from fits to Pantheon+, Union3 and DES-SN5YR data are shown in blue, pink and orange, respectively. Clockwise from the upper left panel, we show the results at maximum crossing function order of $0$, $1$ and $2$.}
\label{U3IScont}
\end{figure}

We show some example contour plots for the hyperparameters used by the crossing functions in Figure~\ref{U3IScont}. Once again, we use the same definition of standard hyperparameter values, equivalent to no deformation, and we show only Union3 template function results, up to $2^{nd}$ order. The overall behaviour, which shows a shift in the Union3 contours to center around the standard hyperparameter values, is reproduced in the contour plots for the other template function, which we do not show here. The remainder of the Union3 template function contour plots may be found in Appendix~\ref{cont}, along with a result summary in Table~\ref{IScontsum}.

It must be borne in mind that the iteratively-smoothed functions from one dataset will be generated in such a way as to match that particular dataset more closely than its own model best fit. Consistency between datasets in spite of the resulting features beyond the $\Lambda$CDM model, which arise solely from the data in a model-independent fashion, would imply mutual consistency under some underlying, extended model.

With 2 free parameters (i.e. $(\mu^{TF}(z)+M_B)\times C^0_0$), both the Union3 and Pantheon+ template functions $(\mu^{TF}$) are only \wlm{compatible}, to any great degree, with their own data. With at least 3 free parameters ($M_B$, $C^{1}_{i=1,2}$), the Pantheon+ template function is able to be \wlm{compatible} with standard hyperparameter values when fit to Union3 and Pantheon+ data. 

In contrast, the Union3 template function requires at least $2^{nd}$ order in crossing functions for \wlm{compatibility} with Union3 and Pantheon+ data. This would seem to imply that the Pantheon+ template function is slightly easier to make \wlm{compatible} than the Union3 template function\footnote{This is similar to what we see in the best fit template function case.}, but it should be noted that both require up to order 4 in Crossing Statistics before they can achieve a $\Delta\chi^2$ larger than the other dataset's own iteratively-smoothed function\footnote{Though the Pantheon+ template is actually fairly comparable already by second order}.

The DES-SN5YR data once again prefers deformations to the Pantheon+ and Union3 template functions for $2^{nd}$ and $3^{rd}$ orders. Of course, when its own iteratively-smoothed function is used as a template to its own data, it is self-consistent at all orders\footnote{This is independent of the chosen $z_{max}$ in the Chebyshev polynomials, see Section~\ref{lowzmax}.}. This would seem to imply that the underlying model preferred by the other two datasets, even outside of $\Lambda$CDM, is less preferred by the DES-SN5YR data.

If we consider the improvement to the $\chi^2$ in this context, in the case of DES-SN5YR data, both Union3 and Pantheon+ template functions require at least $3^{rd}$ order in crossing functions to improve over the intrinsic template function. 

When both these template functions are fit to Union3 and Pantheon+ data, an even larger ($4^{th}$) order is required, although the deformed Pantheon+ template function reaches close to the same improvement in $\chi^2$ at second order, when fit to Union3 data. This signifies that at least some extra freedom away from the data-driven features in each template function must be undone before the fit can be as good as the undeformed iteratively-smoothed function of the data being fit to.

These results so far seem to indicate that DES-SN5YR data do not exhibit as high a level of agreement with the undeformed template functions of the other datasets, compared to the data from Union3 and Pantheon+. However, as seen in the case of the $\Lambda$CDM model, there still exist regions in the space of hyperparameters where all the data are consistent to within at least $2\sigma$, which indicates that particular deformations of an iterative smoothing extension beyond $\Lambda$CDM can result in realisations compatible with each dataset simultaneously.

One final thing to consider, independent of the type of template function used, \wlm{is the striking difference in the redshift range that it covers, which we have previously argued might lead to spurious effects, even though seemingly equivalent priors are used}. Since the template functions are informed by the data, either in terms of the best fit model parameters, or the iteratively-smoothed function, it is reasonable to be suspicious of the effect which this has on the consistency test, and whether the difference in redshift range is leading to a spurious measure of inconsistency. We thus perform the analysis a second time, only considering data within the smallest redshift range of the three datasets, i.e. that of DES-SN5YR (see Section~\ref{lowzmax}). While this allows the contours of DES-SN5YR hyperparameters to shrink to sizes comparable with the other two datasets, the same behaviours described above are still seen to a large degree. This indicates that the results showing a preference tending away from the standard hyperparameter values, but still with non-negligible overlap, are not merely a result of the difference in the redshift ranges of the data.
\subsection{Low $z_{max}$ limit}
\label{lowzmax}
In this analysis, we redo the consistency check in an attempt to remove any possible effects of the mismatch in redshift ranges covered by the different datasets, \wlm{in particular, the DES-SN5YR data}. We make an upper redshift cut at $z_{max}\sim1.1$ corresponding to the maximum redshift in the DES-SN5YR dataset. 

We use the \textit{same} template functions as before, but apply the redshift cut to all data when performing the fits and sampling of hyperparameters; the idea being that, if the fits from a larger redshift range are artificially inconsistent with the DES-SN5YR data because it is only at lower redshift, then this artificial inconsistency should also show up in the low-redshift data of the other datasets. 

This redshift cut also allows us to use the iteratively-smoothed DES-SN5YR function to fit to the other datasets, as extension beyond its $z_{max}$ is no longer required. Again, this will remove the possibility of false inconsistency being registered because of a difference in the redshift range used to generate the template functions, and that over which they are fit.

Although the template function best fit parameters and iterative smoothing parameters are the same as used elsewhere in the analysis, we had to repeat the tests to determine the maximum viable order in crossing functions, finding the suitable value to be $2^{nd}$ order. 

Another important change in this analysis is that the $z_{max}$ used in the Chebyshev polynomials is now that of DES-SN5YR ($z_{max}\sim1.1$). This probes whether the inconsistency found previously was due, in part, to the fact that DES-SN5YR data only had constraining power over a fraction of the range spanned by the Chebyshev polynomials. This is most likely what leads to the effective reduction in the contour size compared to the initial analysis.

\begin{figure}[H]
\centering
\includegraphics[scale=0.33]{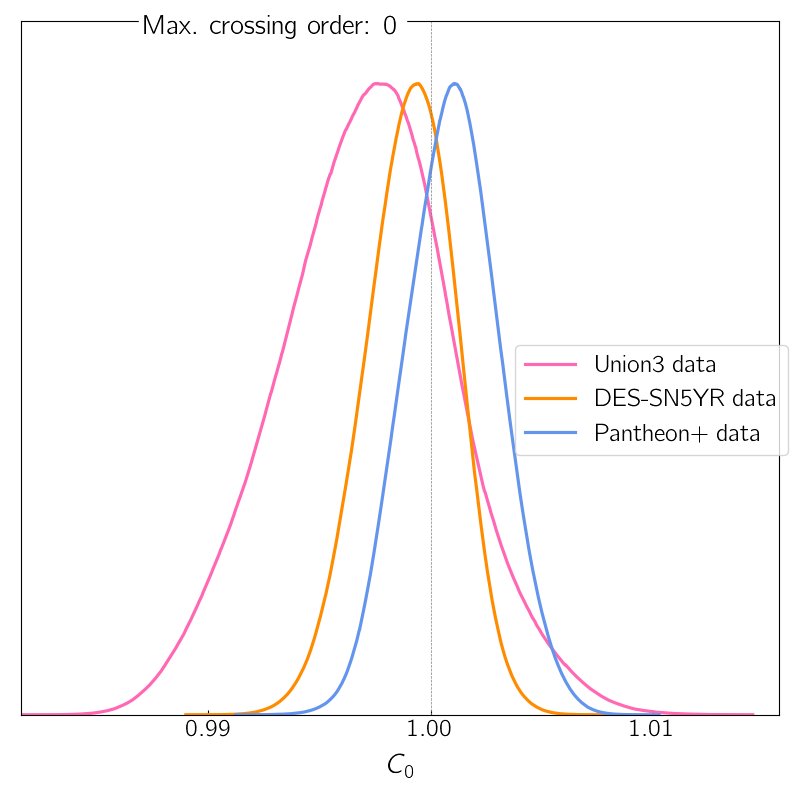}\includegraphics[scale=0.2]{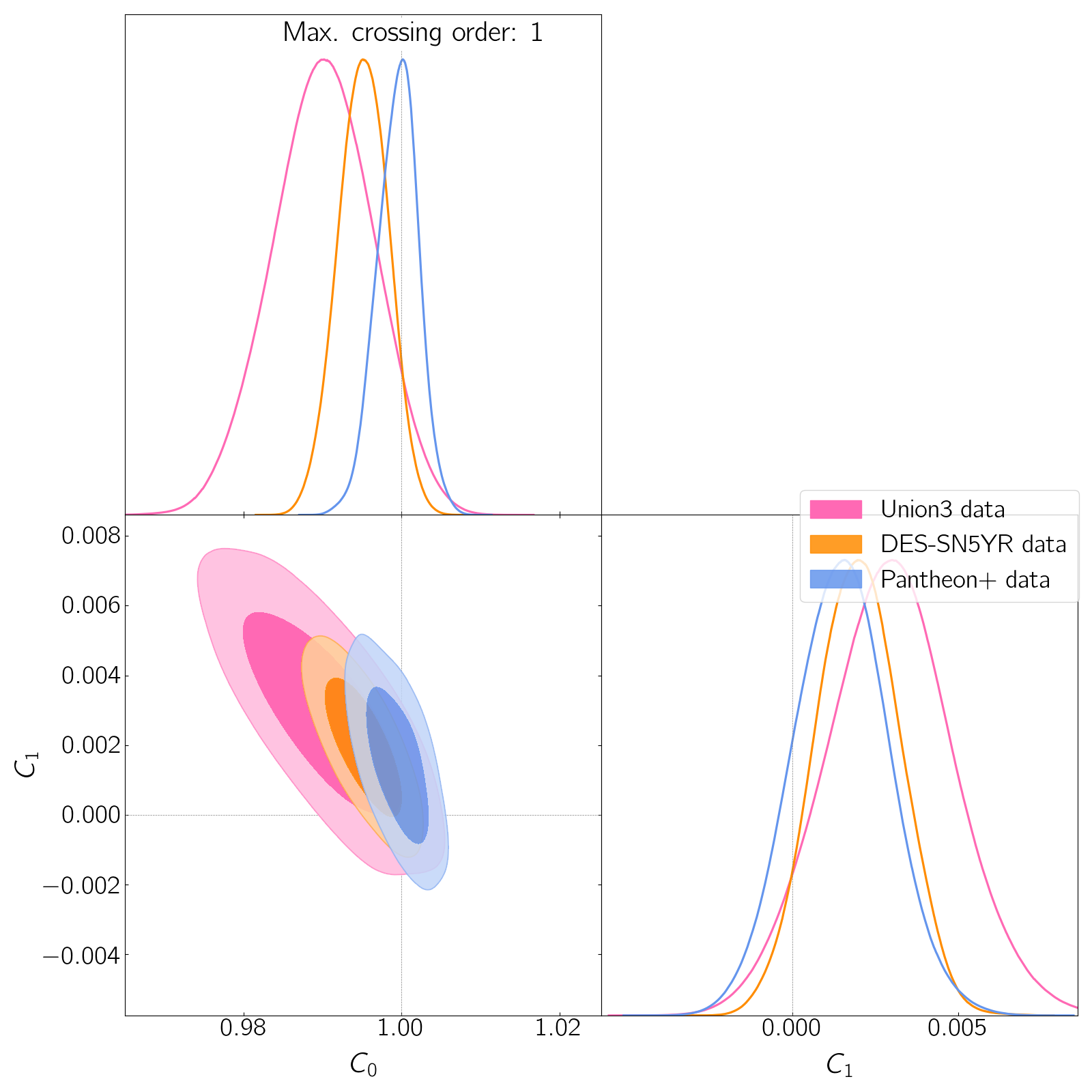}
\includegraphics[scale=0.19]{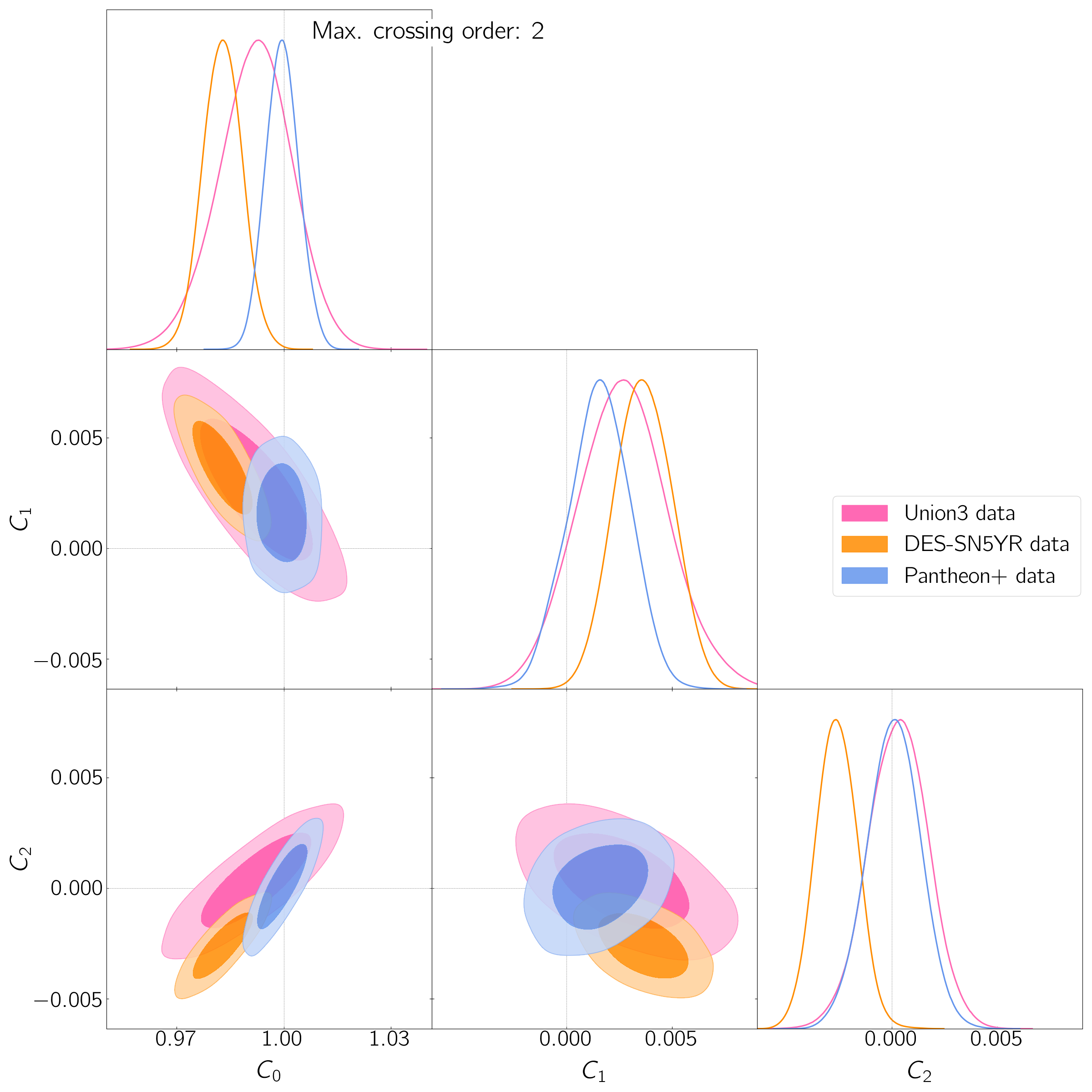}
\caption{Contour plots for the hyperparameters of the \textbf{Union3 $\Lambda$CDM best fit} template function, deformed to fit to all datasets, with the $z_{max}$ cut implemented. The contours from fits to Pantheon+, Union3 and DES-SN5YR data are shown in blue, pink and orange, respectively. Clockwise from the upper left panel, we show the results at maximum crossing function order of $0$, $1$ and $2$. Once again, we see that the portion of the hyperparameter space preferred when fitting the Union3 template function to DES-SN5YR data does not include the standard (undeformed) $\Lambda$CDM case, to at least the $2\sigma$ level.}
\label{U3BFcontlowz}
\end{figure}

\begin{figure}[H]
\centering
\includegraphics[scale=0.33]{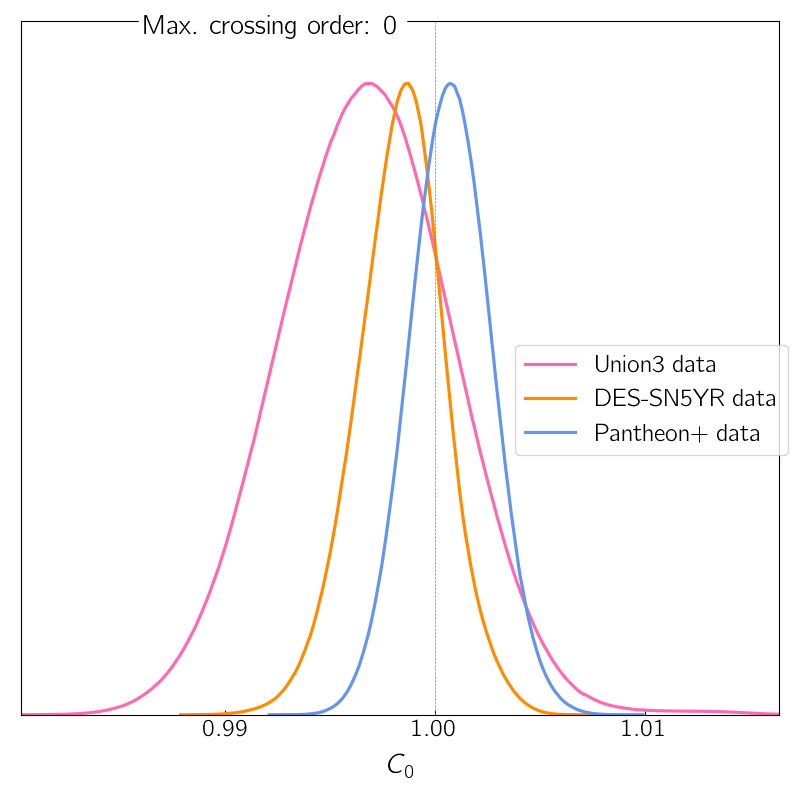}\includegraphics[scale=0.2]{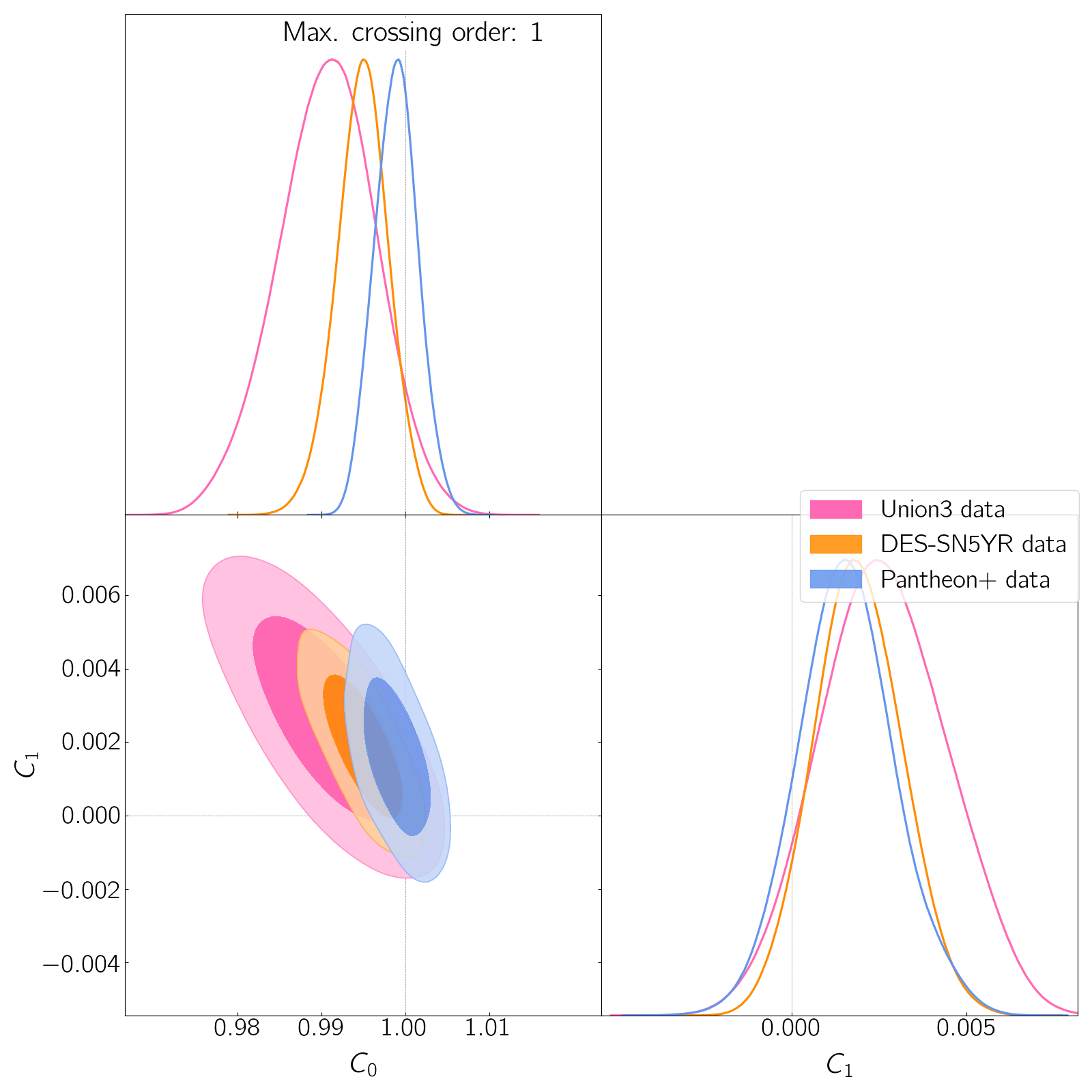}
\includegraphics[scale=0.19]{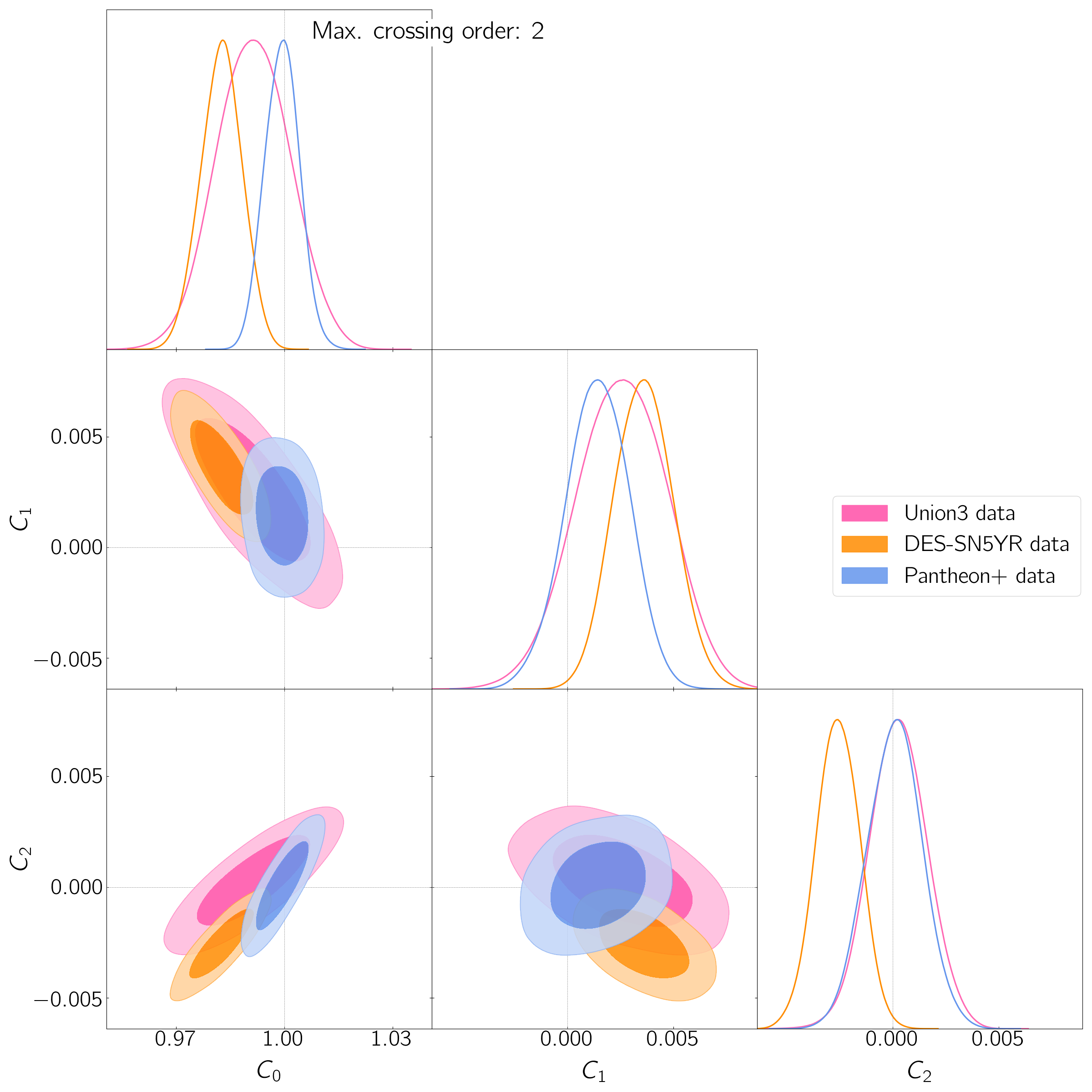}
\caption{Contour plots for the hyperparameters of the \textbf{DES-SN5YR $\Lambda$CDM best fit} template function, deformed to fit to all datasets, with the $z_{max}$ cut implemented. The contours from fits to Pantheon+, Union3 and DES-SN5YR data are shown in blue, pink and orange, respectively. Clockwise from the upper left panel, we show the results at maximum crossing function order of $0$, $1$ and $2$. Here too, we see that even the DES-SN5YR template function, deformed over the reduced range in redshift, prefers a region of the hyperparameter space that excludes the best fit $\Lambda$CDM of the DES-SN5YR data itself, to at least the $2\sigma$ level.}
\label{D5BFcontlowz}
\end{figure}

\begin{figure}
\centering
\includegraphics[scale=0.33]{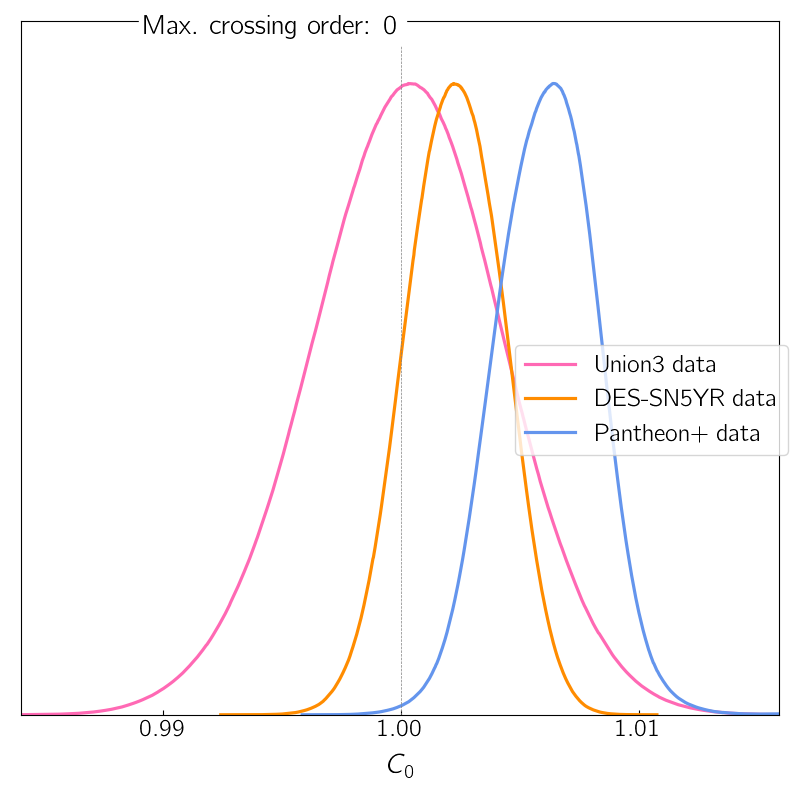}\includegraphics[scale=0.19]{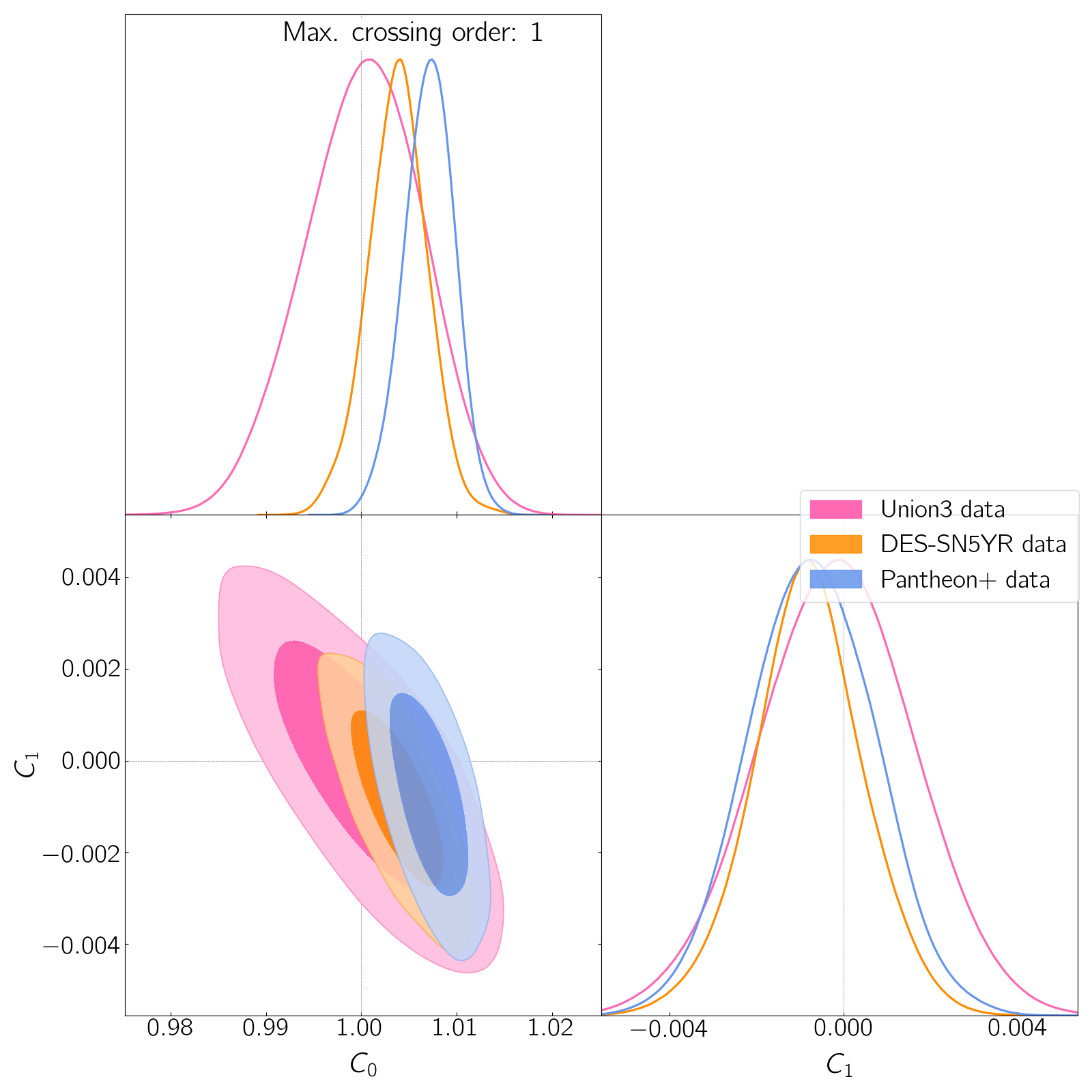}
\includegraphics[scale=0.18]{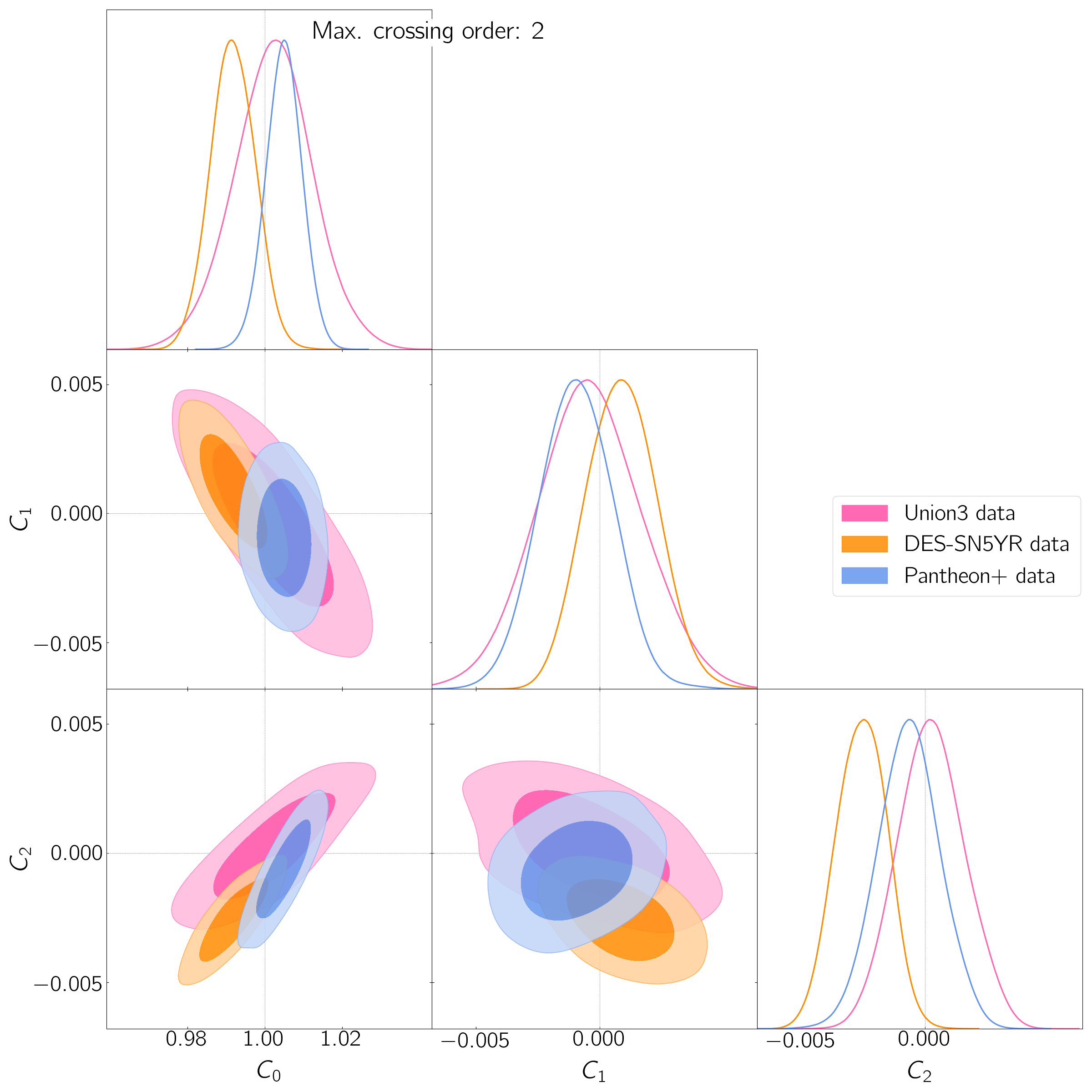}
\caption{Contour plots for the hyperparameters of the \textbf{Union3 iterative smoothing} template function, deformed to fit to all datasets, with the $z_{max}$ cut implemented. The contours from fits to Pantheon+, Union3 and DES-SN5YR data are shown in blue, pink and orange, respectively. Clockwise from the upper left panel, we show the results at maximum crossing function order of $0$, $1$ and $2$.
\vspace{10pt}}
\label{U3IScontlowz}
\end{figure}

\begin{figure}
\centering
\includegraphics[scale=0.33]{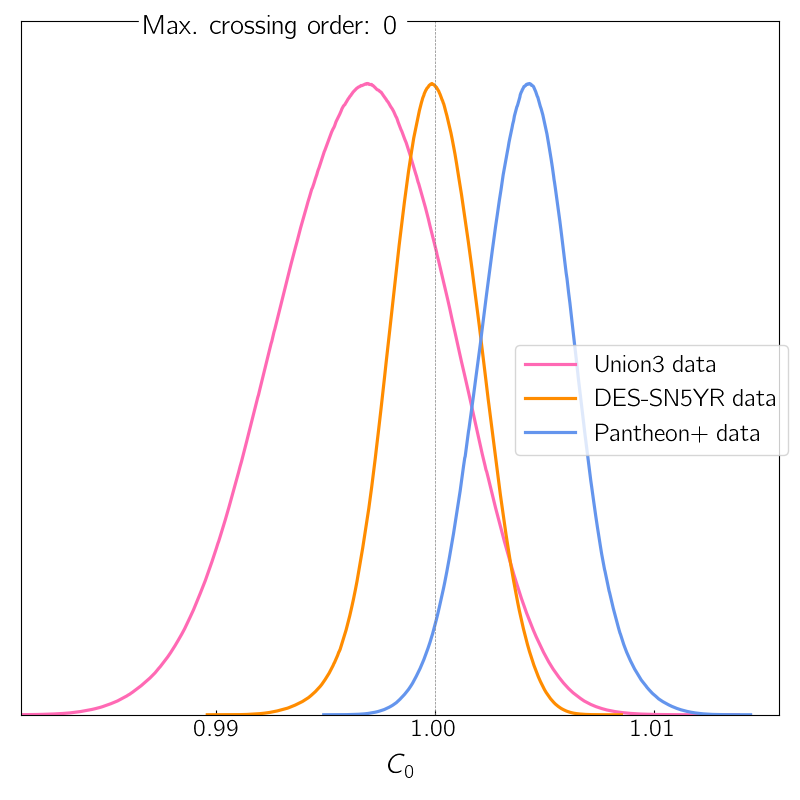}\includegraphics[scale=0.19]{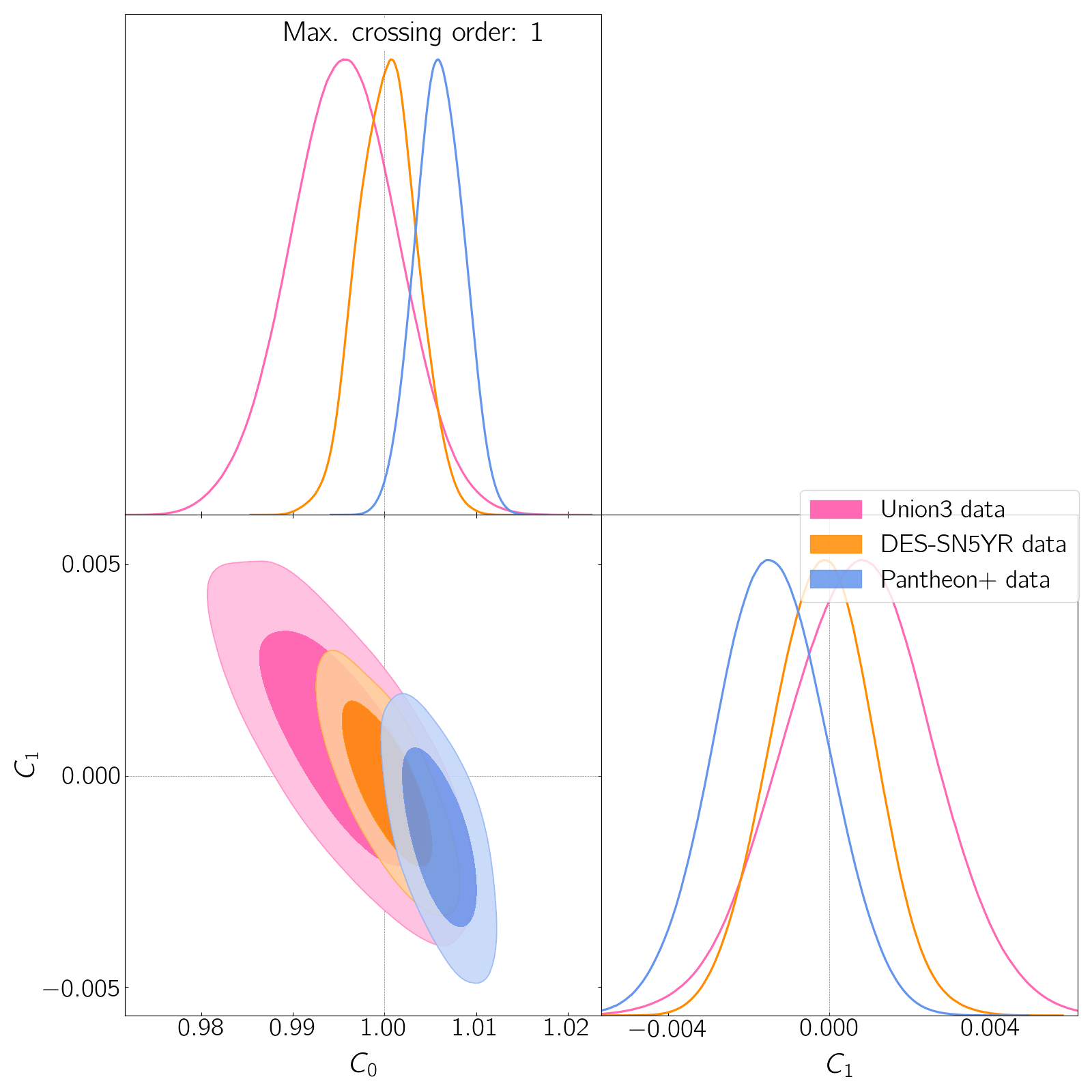}
\includegraphics[scale=0.18]{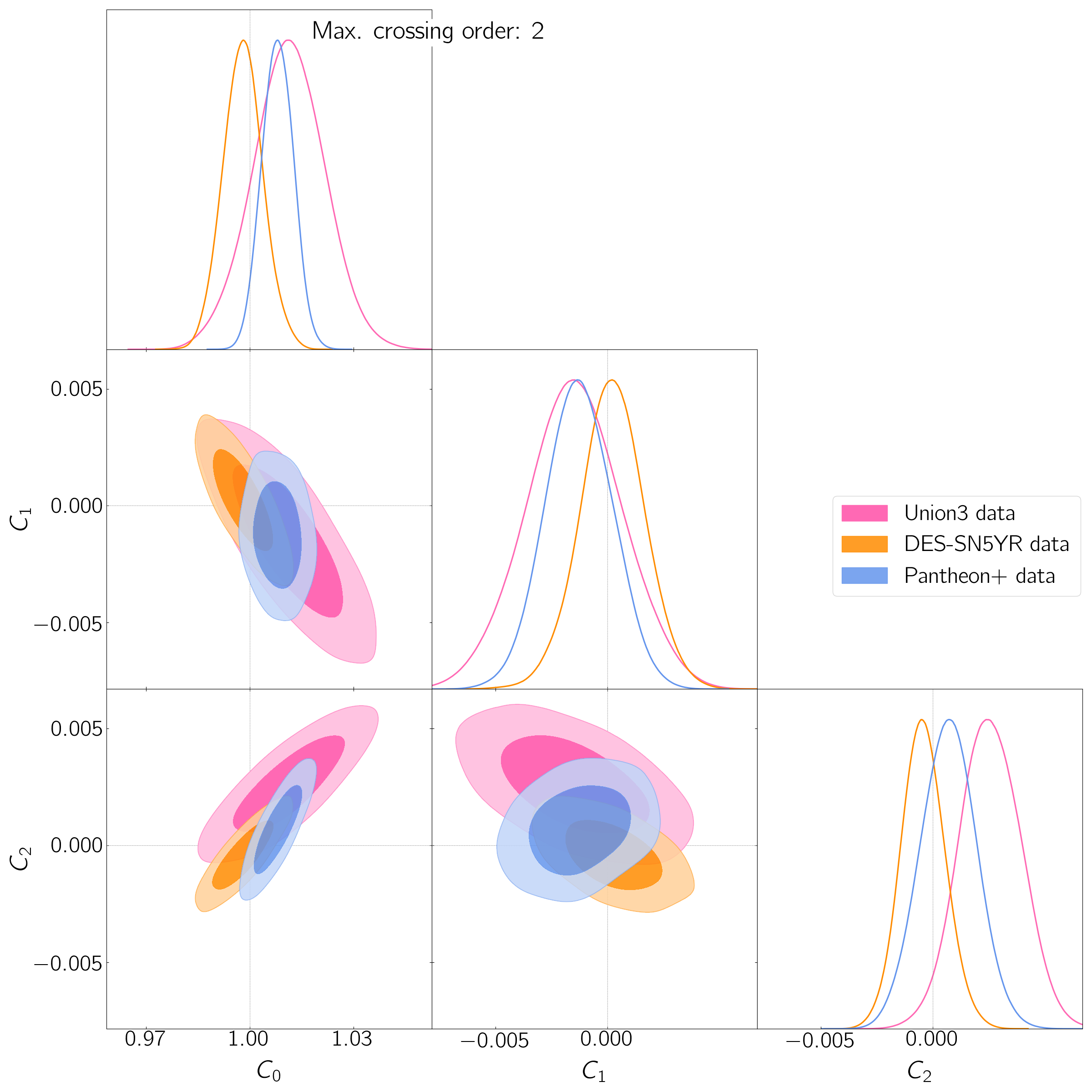}
\caption{Contour plots for the hyperparameters of the \textbf{DES-SN5YR iterative smoothing} template function, deformed to fit to all datasets, with the $z_{max}$ cut implemented. The contours from fits to Pantheon+, Union3 and DES-SN5YR data are shown in blue, pink and orange, respectively. Clockwise from the upper left panel, we show the results at maximum crossing function order of $0$, $1$ and $2$.\vspace{10pt}}
\label{D5IScontlowz}
\end{figure}

In the following subsections, we report the contours plots for the $0^{th}$ to $2^{nd}$ order deformations of the Union3 and DES-SN5YR template functions from Flat $\Lambda$CDM and iterative smoothing (see Figs~\ref{U3BFcontlowz},~\ref{D5BFcontlowz} and~\ref{U3IScontlowz},~\ref{D5IScontlowz}), as examples from these two most recent, and distinct compilations. The behaviour of the Pantheon+ results are very similar (see Tables~\ref{lowzBFcontsum} and~\ref{lowzIScontsum} for a summary).

\subsubsection{$\Lambda$CDM best fit template function}

In general we see that the contours of Union3 and Pantheon+ are very slightly larger than before, which is to be expected since we are fitting to a subset of the data with template functions produced by the entire data range. It is also clear that the DES-SN5YR contours are tighter, and that the best fit value of the hyperparameters shift closer to the standard values. The combination of these effects results in much the same behaviour as we have seen previously (compare the first three lines of Tables \ref{BFcontsum} and \ref{lowzBFcontsum}): The hyperparameters of Union3 and Pantheon+ are routinely between $1-2\sigma$ of the standard values, while DES is once again slightly further out of agreement (though less so than in the full redshift analysis). 

The only practical difference in the latter case is that instead of deforming the previous best fit template functions from each dataset to fit the DES-SN5YR data covering a portion (about half) of the rescaled $\tilde{z}$ range, the deformations are fixed by the same data spread over the entire range of $\tilde{z} \in [-1,1]$. This means that deformations of a particular order now occur at a smaller scale, relative to the extent of the data. The hyperparameter values thus no longer have to be as far from standard values \wlm{to achieve a model closer} to the data (as compared to when the data only covered a part of the $\tilde{z}$ range).

The change in the range of the data relative to the effective Chebyshev polynomial range is also the reason why the degeneracy direction of the contours is prone to change, compared 

\hspace{-25pt}to the full redshift analysis. Now that the preferred shape of the deformations is governed by a subset of the original data, rescaled to fit into the interval $[-1,1]$, the behaviour of the contours may change. The exact effect on the contours is difficult to predict, in general.

We also observe that there is once again a region of overlap of all 3 datasets' contours within the space of hyperparameters, indicating that some subset of deformations can be responsible for producing each one concurrently, to within $1\sigma$ or $2\sigma$, depending on the maximum order of crossing functions considered.

\begin{table}[H]
\small
\centering
\begin{tabular}{c|c c c|c c c|c c c}
Flat $\Lambda$CDM T.F.: &\multicolumn{3}{c|}{Pantheon+}&\multicolumn{3}{c|}{Union3}& \multicolumn{3}{c}{DES-SN5YR} \\
Data: & P+ & U3  & D5& P+ & U3  & D5& P+ & U3  & D5\\
\hline
T.F. (with $M_B$)&1391.42&23.1294&1641.61&1392.24&22.1510&1640.16&1391.88&22.1983&1640.09\\
$2^{nd}$ order&1390.72&19.1975& 1630.59&1390.84&19.2010&1630.58&1390.80&19.2005&1630.55
\end{tabular}
\caption{$\chi^2$ achieved from $2^{nd}$ order crossing functions applied to the template functions from the Flat $\Lambda$CDM best fit, compared to fitting just a translational factor $M_B$.}
\label{lowzBFchi2}
\end{table}

\subsubsection{Iterative smoothing template function}

In the case of iteratively-smoothed template functions, the results from the analysis including the lower maximum redshift cut are similar to the full redshift analysis. The Union3 and Pantheon+ contours show about the same level of mutual consistency as found when considering the full redshift range. 

As in the case of the best fit template functions with low $z_{max}$ cut-off, we observe that the DES-SN5YR contours have shrunk, due to the improved constraining power of the data, which now extends over the entire range of relevant redshifts. The iteratively-smoothed template functions once again conform very well to the data used to produce them (see Table \ref{lowzIScontsum}, in comparison to Table~\ref{IScontsum}). As for the mutual agreement of the three datasets, the overall conclusions are much the same as in the full redshift analysis, save for the DES-SN5YR contours whose combined tightening and movement towards the other contours make them, on the one hand, closer to the standard hyperparameter values slightly more frequently and, on the other, overlap with lower $\sigma$ regions of the other contours more often than before. 

\begin{table}[H]
\footnotesize
\centering
\begin{tabular}{c|c c c|c c c|c c c}
Iterative Smoothing T.F.: &\multicolumn{3}{c|}{Pantheon+}&\multicolumn{3}{c|}{Union3}& \multicolumn{3}{c}{DES-SN5YR} \\
Data: & P+ & U3  & D5& P+ & U3  & D5& P+ & U3  & D5\\
\hline
T.F. (with $M_B$)&1389.77&22.2137&1644.79&1404.39&19.0586&1637.66&1405.45&23.4657&1629.03\\
$2^{nd}$ order&1389.19&19.1414&1631.51&1394.80&18.9826&1629.46&1399.62 &19.2877&1628.85
\end{tabular}
\caption{$\chi^2$ achieved from $2^{nd}$ order crossing functions applied to the template functions from iterative smoothing, compared to using just a translational factor $M_B$.}
\label{lowzISchi2}
\end{table}

\vspace{-10pt}
\section{Conclusion}
\label{conc}
We have used Crossing Statistics and Iterative Smoothing to assess the mutual consistency of three supernova datasets in a model-independent manner. Given the characteristics of the data compilations, we have performed this analysis for both the largest and smallest redshift ranges over which the three datasets extend. The choice of maximum $z$ as the overall maximum redshift of all three datasets was seen to have a small effect on results, moving the crossing hyperparameters slightly further from their standard values, thus suggesting that the DES-SN5YR data prefer slightly larger deformations than otherwise would be the case, as well as increasing the size of the contours (since the range of DES-SN5YR data did not cover the constrained range of redshifts as well as the other two datasets could). Nevertheless, the conclusions that may be drawn from both analyses are very similar, and are reported here.

In the case of the first null hypothesis, relating to mutual consistency given the best fit to the assumed flat $\Lambda$CDM model, we may say that it is rejected only at slightly above $2\sigma$ level. This in the case of DES-SN5YR data and the undeformed best fit template functions from either of the two other datasets; whereas, those datasets show a higher level of mutual consistency to each other's standard hyperparameter values (i.e. undeformed best fits). However, all 3 datasets share enough regions within the extended space of hyperparameters that they may all be generated by the same subset of deformations, to within $1-2\sigma$. 

In the case of iterative smoothing template functions, where we are interested in mutual consistency under the implicit model that provides a \wlm{closer} fit to a particular dataset, the data prefer deformations that are slightly more extreme with respect to the template function, i.e. their contours are, in general, further from standard hyperparameter values. We see here too that DES-SN5YR prefers hyperparameter values that are slightly less \wlm{compatible} with the standard values, as compared to Union3 and Pantheon+ to each other, though this is more pronounced in the full redshift analysis than with the low $z_{max}$ limit. Thus, the overall results suggest that there is some disagreement of the DES-SN5YR data with the model-independent smooth functions through the other data. Since this represents an underlying model beyond flat $\Lambda$CDM, we conclude that it rejects the part of second null hypothesis, which concerns the mutual consistency of the datasets with the model-independent template functions of the others, at around 2$\sigma$. Despite this, all three datasets show at least some regions of mutual consistency in the spaces of the hyperparameters controlling deformations around all of the the available template functions. This overlap, which is more pronounced in the low $z_{max}$ analysis, indicates that the data are all still able to be produced by a subset of the hyperparameter space of extended models, and thus consistent to within $1-2\sigma$ ($\sim1\sigma$ in the low $z_{max}$ analysis), in that sense.

It would be interesting to explore the kind of expansion histories suggested by each the datasets, and their possible theoretical explanation and implications, as well as the volume, in the space of hyperparameters, which corresponds to the subset of deformations to the template functions which are preferred by all of the datasets. This is left to future works.

Overall, within the framework of our methodology, the three datasets show a good level of mutual consistency, though their preferred hyperparameters do not generally coincide with each other's best fit $\Lambda$CDM models. In the case of a model-independent template function, with \wlm{closer} fit to each dataset's particular features, this remains the case. However, within the implied model space provided by the deformations from crossing functions, there are always regions, with $1-2\sigma$ agreement in the hyperparameter values, to be found. Thus, the datasets can be said to be mutually consistent at this level, though not necessarily within that portion of the hyperparameter space corresponding to the concordance cosmology. This means that it is at least possible that no individual dataset is solely systematically incorrect, and points towards the possibility of interesting future studies where the corresponding cosmological implications can be explored.
\section{Acknowledgements}
The authors would like to thank Rodrigo Calderon and Wuhyun Sohn for useful discussions during the preparation of this work.
\begin{appendices}

\section{Tuning the hyperparameters}
\label{tuning}
\vspace{-5pt}
\subsection{Crossing functions}
Aside from the deformed function being a good fit to the data, we must choose a maximum order in Crossing Statistics at which to truncate the series.
Here we choose a maximum polynomial order of 4 for the crossing functions, based on the following considerations:

\textbf{Physicality:} The distance modulus should increase monotonically with redshift. If, at a particular order in Crossing Statistics, the function is no longer monotonic, we should not include that order. We also impose that the Hubble factor should not decrease with increasing redshift. 
\vspace{-15pt}
\begin{table}[H]
\centering
\begin{tabular}{c|c c c}
\multirow{2}{*}{Template function} & \multicolumn{3}{c}{Data}\\
\cline{2-4}
 & Pantheon+ & Union3  & DES-SN5YR\\
\hline
Pantheon+ & $\geq6$ (4)& 4 (5)& $\geq6$ ($\geq6$) \\
Union3  & $\geq6$ (5)& 4 (4) &  $\geq6$ ($\geq6$)\\
DES-SN5YR &$\geq6$ (-)& $4$ (-)& $\geq6$ ($\geq6$)\\ 
\end{tabular}
\vspace{-5pt}
\caption{Maximum orders in Crossing Statistics at which the monotonicity of the distance modulus and positive gradient of $H(z)$ are preserved. We consider the template functions from the Flat $\Lambda$CDM best fit (and the iterative smoothing function)}
\label{XingMonot}
\end{table}

\textbf{Small fluctuations:} We consider only small deformations around the chosen template function, at the level of $\sim3\%$. 
\vspace{-15pt}
\begin{table}[H]
\centering
\begin{tabular}{c|c c c}
\multirow{2}{*}{Template function} & \multicolumn{3}{c}{Data}\\
\cline{2-4}
 & Pantheon+ & Union3  & DES-SN5YR\\
\hline
Pantheon+ & 5,6 (5,6) & 6 (3,4,6)& $2$ ($2,5$)\\
Union3  & 5,6 (5)& 6 (3,4,6) &  $2$ ($2,5$)\\
DES-SN5YR & 5,6 (-) & 3,4,5,6 (-) & $2$ ($2,5$)\\ 
\end{tabular}
\caption{Orders in Crossing Statistics ($\leq 6$) at which the deformation is $\geq 2\%$ of the template function at at least one redshift. We consider the template functions from the Flat $\Lambda$CDM best fit (and the iterative smoothing function)}
\label{XingSmall}
\end{table}
Note: Consistency requires that we compare the different datasets at the same order in crossing, so we can only do this at the highest order at which the above criteria are met, i.e. $n=4$. There are some cases, particularly at lower orders, where the fits to Union3 and DES-SN5YR data are deformed by slightly more than $2\%$. Taking this into account, and the maximum order $n=4$ from physical arguments, in all cases finally considered, the magnitude of deformation relative to the template function never exceeds $2.5\%$.

\wlmn{\textbf{Degrees of Freedom:} We are interested in extended models with additional parameters that are actually justifiable. In some limit of increasing of crossing parameters, the deformed function will eventually have enough freedom for all datasets to be consistent, at the expense of inflated contours (see, for example, the lower panels of Figs~\ref{U3BFcont34} and~\ref{D5BFcont34}). While this isn't necessarily guaranteed to happen within the maximum order allowed by the above criteria, it should be checked. In order that we actually consider some deformed template function to be a good fit to the data, it is necessary that we do not include superfluous parameters, thereby increasing the degrees of freedom without correspondingly achieving an appreciable decrease in the $\chi^2$. Otherwise ``consistency'', as we define it, could be trivially obtained, by deliberately enlarging the hyperparameter contours. Thus, we look for extended models with only relatively few additional parameters that show a significant improvement in $\chi^2$.}

If we assume that the template function is the true underlying model of the data, then the $\Delta\chi^2$ of the deformations with respect to the template function will itself follow a chi-square distribution with number of degrees of freedom equal to the number of parameters fit to produce the deformed function.\footnote{Technically this is only true for a sufficiently large dataset \cite{wilks}. In the case of Union3, this might require further investigation; however, given the good agreement with Pantheon+, we do not think it necessary in this work.} \wlmn{We may then use the cumulative distribution functions (CDFs), assuming a chi-square distribution with the requisite number of degrees of freedom \cite{NumRec}, to calculate the probability that a random draw would produce a given $\Delta\chi^2$ value.}

\vspace{-10pt}
\begin{figure}[h!]
\hspace{-45pt}\includegraphics[scale=0.3]{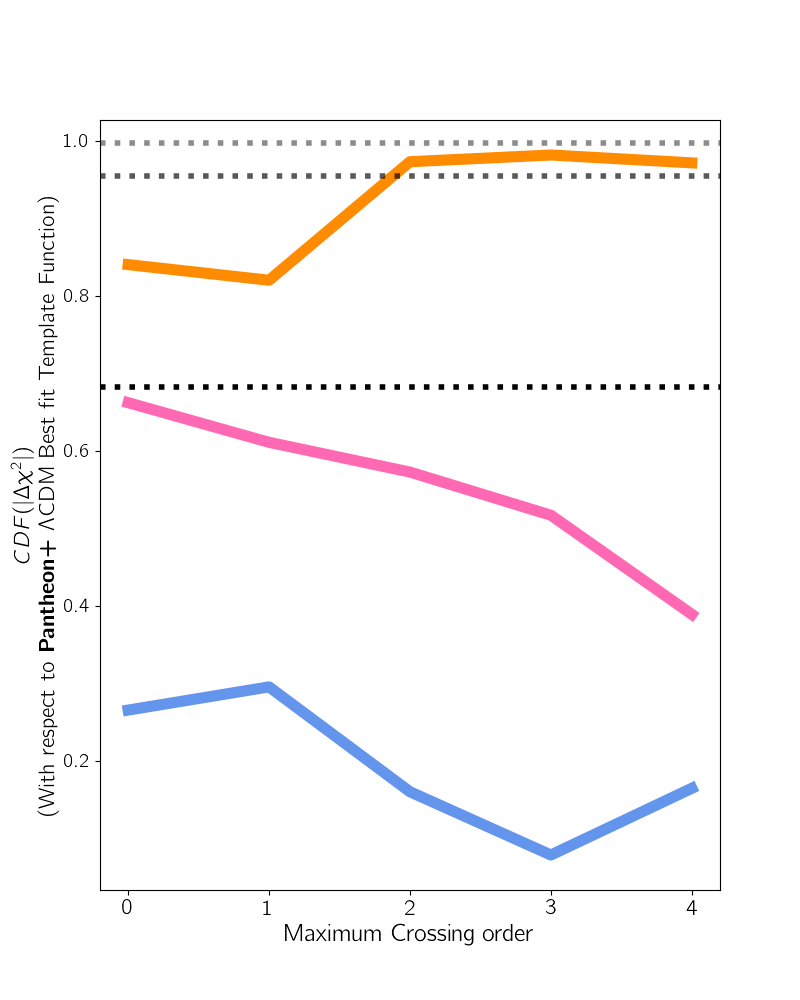}\includegraphics[scale=0.3]{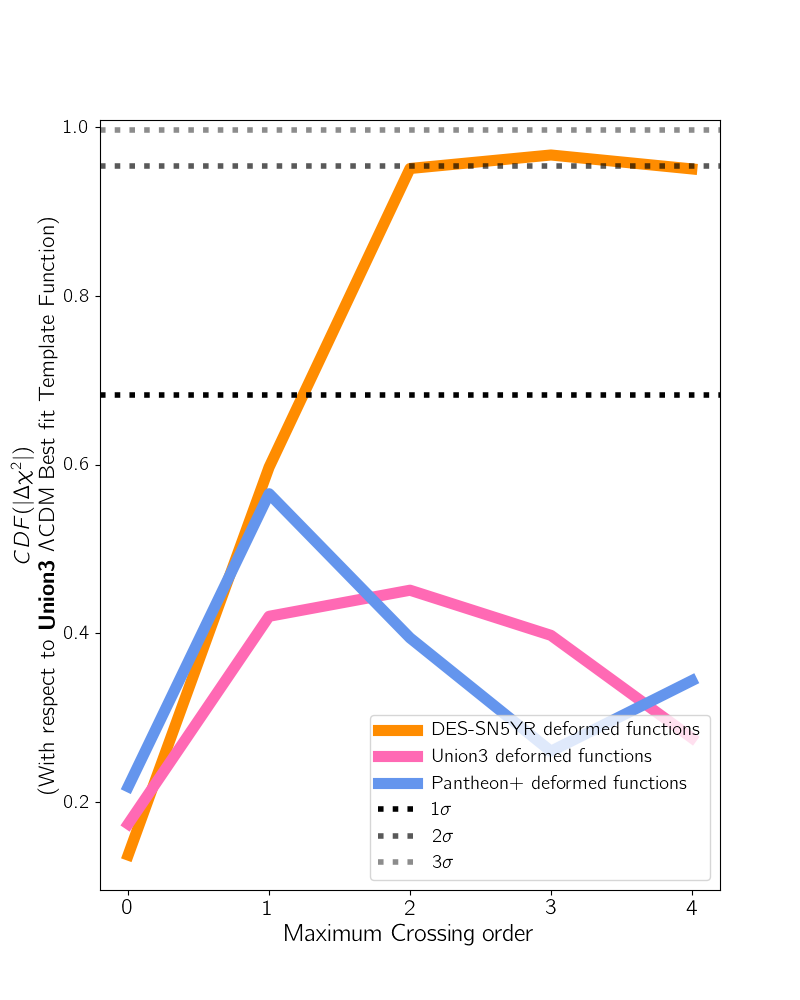}\includegraphics[scale=0.3]{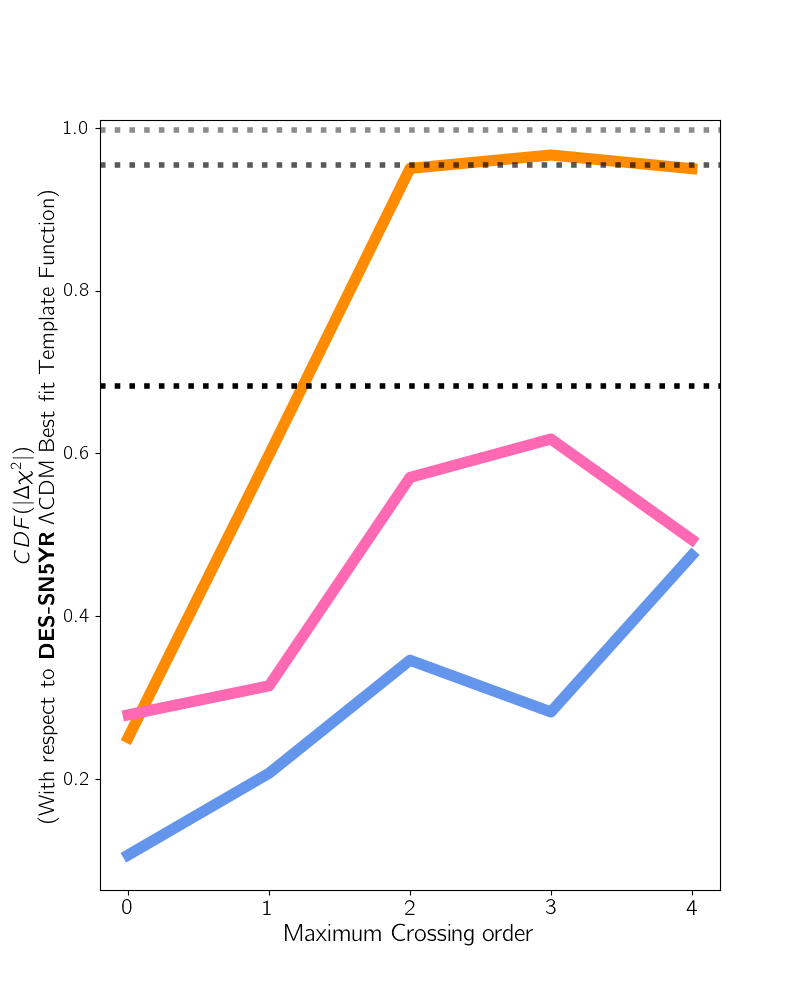}
\caption{\wlmn{The cumulative distribution functions (CDFs) for the $\Delta\chi^2$ obtained by each deformation of the template function are shown in solid lines. Each panel signifies a different $\Lambda$CDM best fit template function (\textit{Left panel}: Pantheon+, \textit{Middle panel}: Union3, \textit{Right panel}: DES-SN5YR), while the colour convention is used to show which dataset this template function is deformed to fit. In other words, $1-CDF$ in each plot and for each Crossing order corresponds to the cumulative probability of the given data assuming the template function.}}
\label{CDF}
\end{figure}

\wlmn{In Fig.~\ref{CDF}, we see these probabilities for each template function, deformed to each dataset. The curves correspond to Fig.~\ref{n0dchi}, and may be interpreted in terms of the number of $\sigma$ confidence at which the $\Delta\chi^2$ of the deformed function is distinct from that of the particular template function, given some data. The feature of interest is the point at which the decrease in $\chi^2$, compared to that of the template function, is no longer sufficient to improve the probability, given the number of additional free parameters. In the full redshift range case, we see that adding higher orders of crossing hyperparameters is still useful in at least one set of deformations for each template function, until at most $4^{th}$ order. For this reason we can reasonably consider the results from Union3 and Pantheon+ up to maximum crossing order of $2$ in the full redshift analysis.} 

\wlmn{In the case of the reduced redshift range, where DES-SN5YR may be fairly considered, our prior criteria already exclude analysis above $2^{nd}$ order in crossing statistics, and the significance of the corresponding $\Delta\chi^2$ in this range is just as high, if not more so.}

\subsection{Iterative smoothing}
It is also necessary to choose values for $\Delta$ and $N$ to be used in the iterative smoothing of each dataset. To reiterate, we aim to obtain a smooth function which has a viable fit to the dataset, without fitting the small-scale noise and independent of cosmological parameters.

The choice of $\Delta$ controls the sensitivity to the neighbouring data points when determining the value of the iterative smoothing function at a particular point. The precise details of the choice of $\Delta$ depend on the noise characteristics of the data, and the number of supernovae in the redshift range considered. The detailed explanation of previous choices in the literature may be found in~\cite{Shafieloo:2005nd,Shafieloo_2007} with some examples in~\cite{Shafieloo_2010, Koo:2020wro}. In these cases, the goal was reconstruction or cosmological parameter estimation and, as such, the considerations for the choice of $\Delta$ concerned the biasing and error introduced in derived parameters. 

In this case, since we are only interested in mutual consistency and not in reconstruction per se, we have a little more freedom in the choice of $\Delta$\footnote{As explained in Section~\ref{miTF}, the aim of the iteratively-smoothed function is to pick up some of the features of the particular dataset and to be a \wlm{viable fit without using the concordance cosmology parameters}. The effects from using iterative smoothing to generate a template function will be moderated by the Crossing Statistics that are subsequently applied: If there are features inherited by the iterative smoothing template function that are due only to the small-scale noise characteristics of the particular dataset used, and not a common underlying model, then the subsequent versions with smooth deformations from the applied crossing functions will not be able to reconcile the smoothed function as a good fit to the other datasets, without much higher order in crossing functions.}. Thus, after testing various values of smoothing length and the number of iterations required for convergence, in the case of Pantheon+ we adopt a value of $\Delta=0.4$, slightly larger than in the literature, to ensure convergence within a more reasonable number of iterations. The DES-SN5YR dataset has not only slightly more supernovae, but also a much smaller redshift range than Pantheon+ (about half), and so the density of data points is higher. For this reason, we opt for a slightly smaller value of $\Delta = 0.3$. In the initial analysis, we do not use the iteratively-smoothed function from DES-SN5YR as a template function for other datasets, because of the limited redshift range, but we do perform a cross-check using the dataset itself. The Union3 dataset has fewer data points, and so the value of $\Delta$ must be increased in order that each point of the smoothing function be sensitive to at least one datapoint. We use $\Delta=0.6$. 

The choice of $N$ is determined by the convergence in $\Delta\chi^2$. For a given smoothing length, N must be sufficiently high that subsequent iterations do not give any meaningful improvement in the $\chi^2$. In this case we found, at most, $N=2000$ to be a suitable maximum number of iterations over all three datasets (see Fig.~\ref{IS_dchi})\footnote{Even where the convergence happens after fewer than $2000$ iterations, continuing iterating until $N=2000$ will not introduce any issues, since there is no significant change in the resulting function.}.

In addition to these choices of hyperparameters, we discovered some convergence issues when using iterative smoothing on the Union3 dataset. This was found to be due to the relatively low number of \wlm{splined} data points, and a relatively strongly negative off-diagonal element in the inverse of the covariance matrix for the correlation between the $3^{rd}$ and $4^{th}$ \wlm{spline nodes}. The element itself is not the most strongly negative one in the inverse covariance matrix, but the way that the iterative smoothing method operates means that it causes a problem. In the $(k+1)^{th}$ iteration, the previous estimate has the following term added to it:
\be
\frac{\boldsymbol{\delta\mu}_k^T \cdot {\rm C}^{-1} \cdot \mathbf{W}(z)}{\mathbb{1}^T\cdot {\rm C}^{-1} \cdot \mathbf{W}(z)}
\ee
where the numerator is a measure of the residuals present in the previous iteration's estimate, each weighted by the distribution function $W_i(z)$ and the inverse covariance matrix (a measure of the uncertainty/freedom present in the data). The denominator is a normalisation factor, meant to ensure that at each step the contribution from this factor is of the order of the residuals, for convergence. It comprises the weighting function $W_i(z)$ and $\mathbb{1}^T\cdot {\rm C}^{-1}$, which is a vector whose elements are each the sum of one column of the inverse covariance matrix. 

If there are sufficiently many, or sufficiently large, negative elements compared to positive ones in a particular column, then the result of this sum may become very small, while the numerator's equivalent $\boldsymbol{\delta\mu}_k^T \cdot {\rm C}^{-1}$ does not, thus causing the additive factor to become very large, and preventing convergence. Therefore, for the purposes of generating the Union3 iterative smoothing template function, we elected to set the offending off-diagonal element of $C^{-1}$ manually to exactly zero. However, in all calculations of $\chi^2$ and likelihood, we make use of the original, full inverse covariance matrix.

\section{Contour plot summary}
\label{cont}
\begin{figure}[H]
\centering
\includegraphics[scale=0.13]{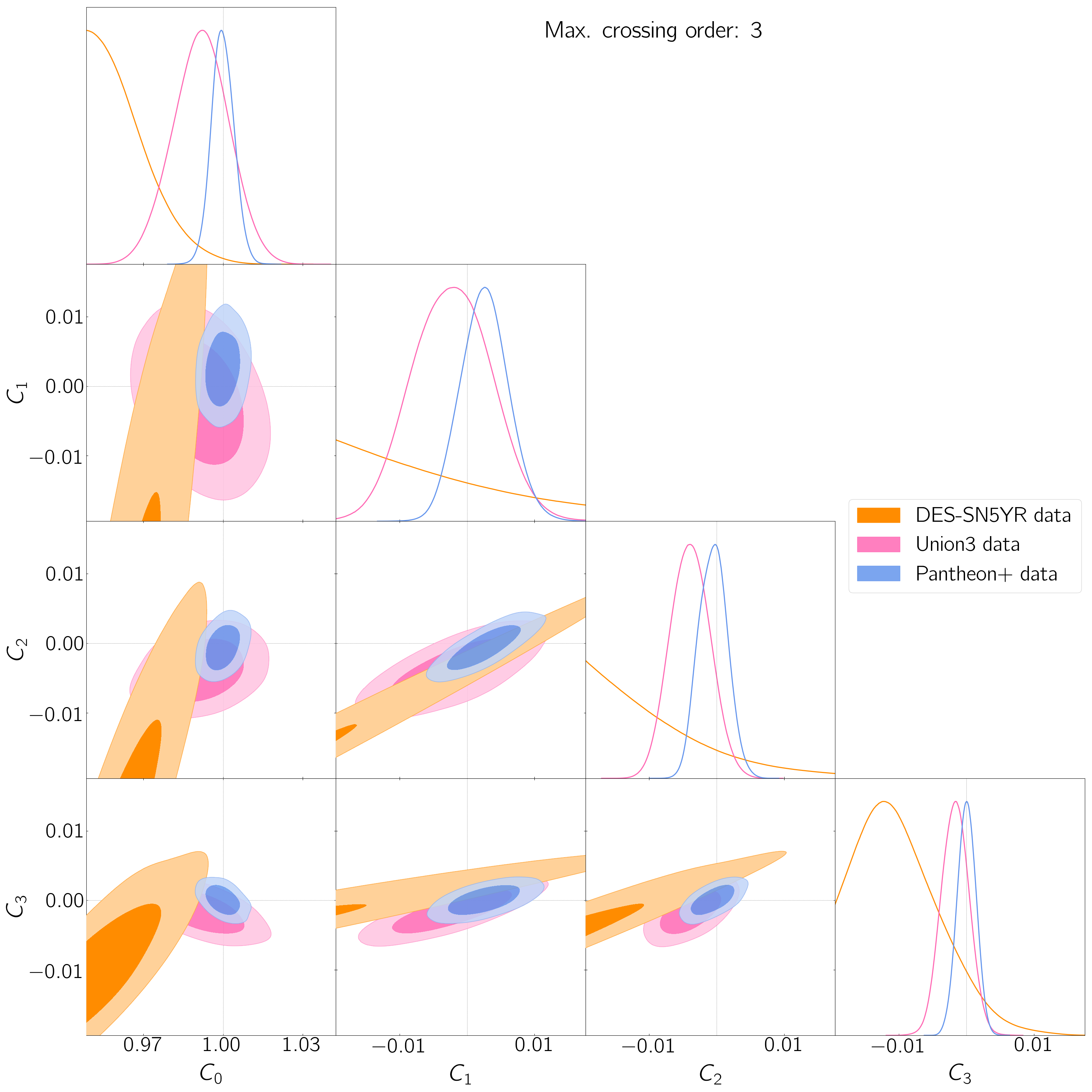}
\includegraphics[scale=0.11]{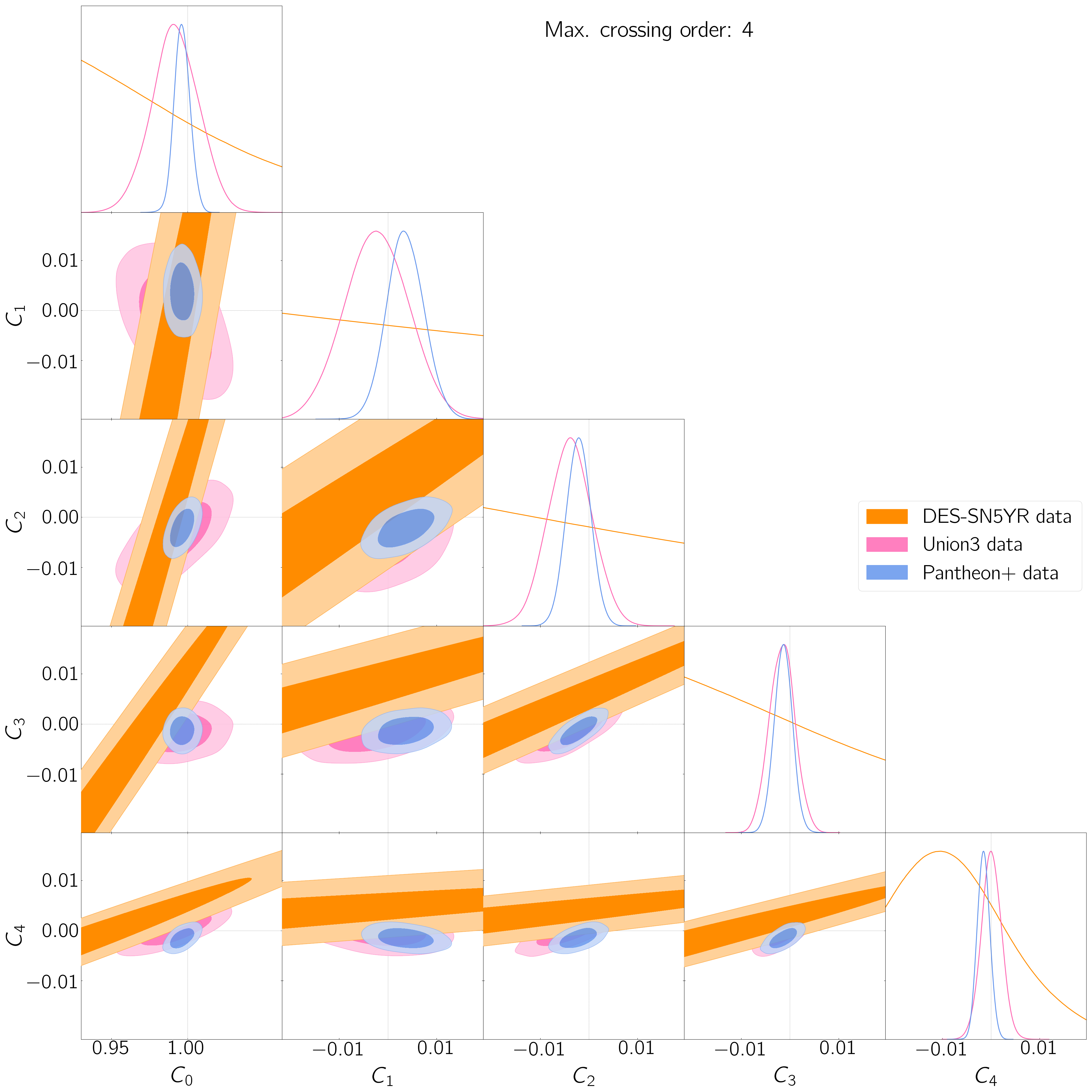}
\caption{Contour plots for the hyperparameters of the \textbf{Union3 $\Lambda$CDM best fit} template function, deformed to fit to all datasets. The contours from fits to Pantheon+, Union3 and DES-SN5YR data are shown in blue, pink and orange, respectively. We show the results at maximum crossing function order of $3$ (upper panel) and $4$ (lower panel), for the full redshift range.}
\label{U3BFcont34}
\end{figure}

\begin{figure}[H]
\centering
\includegraphics[scale=0.13]{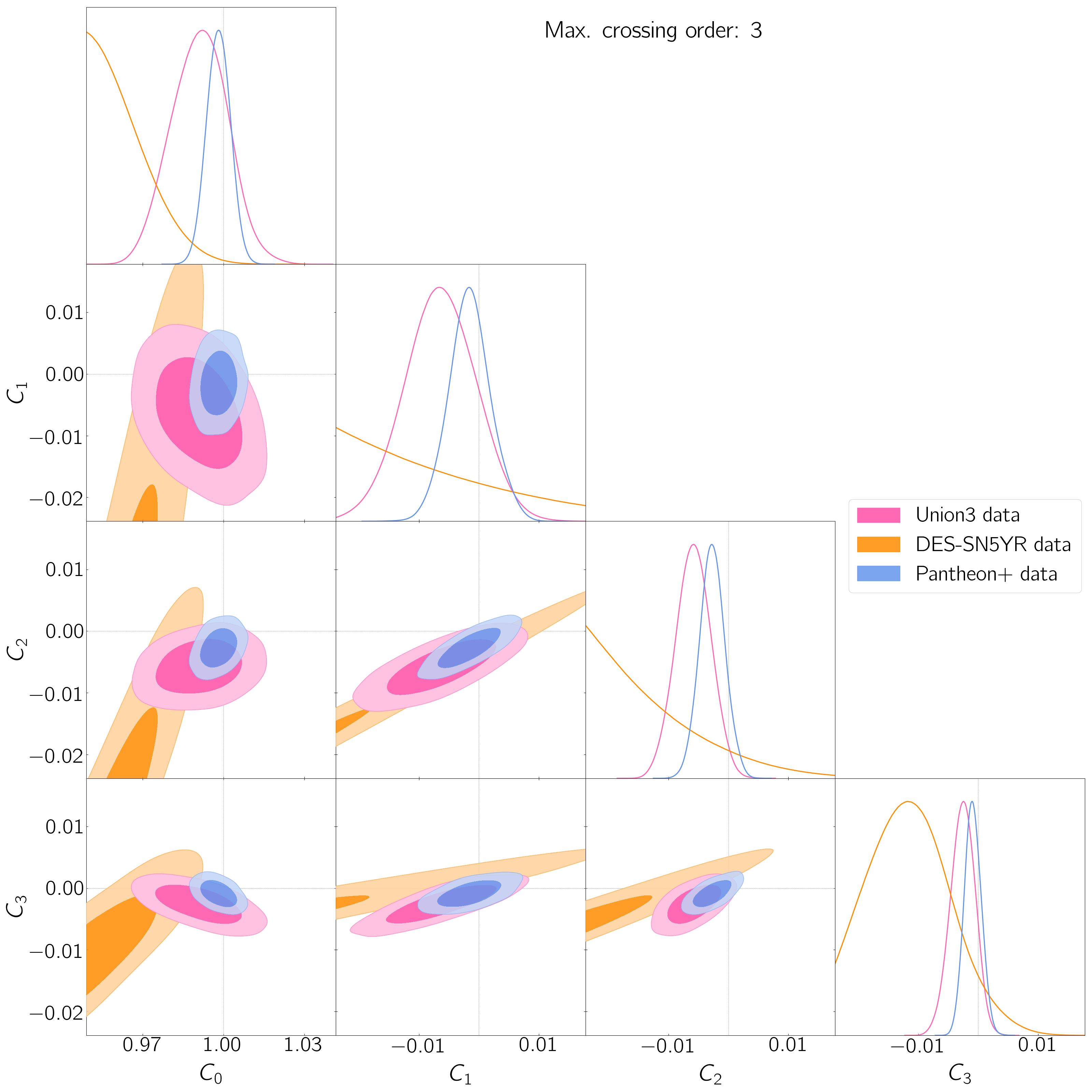}
\includegraphics[scale=0.11]{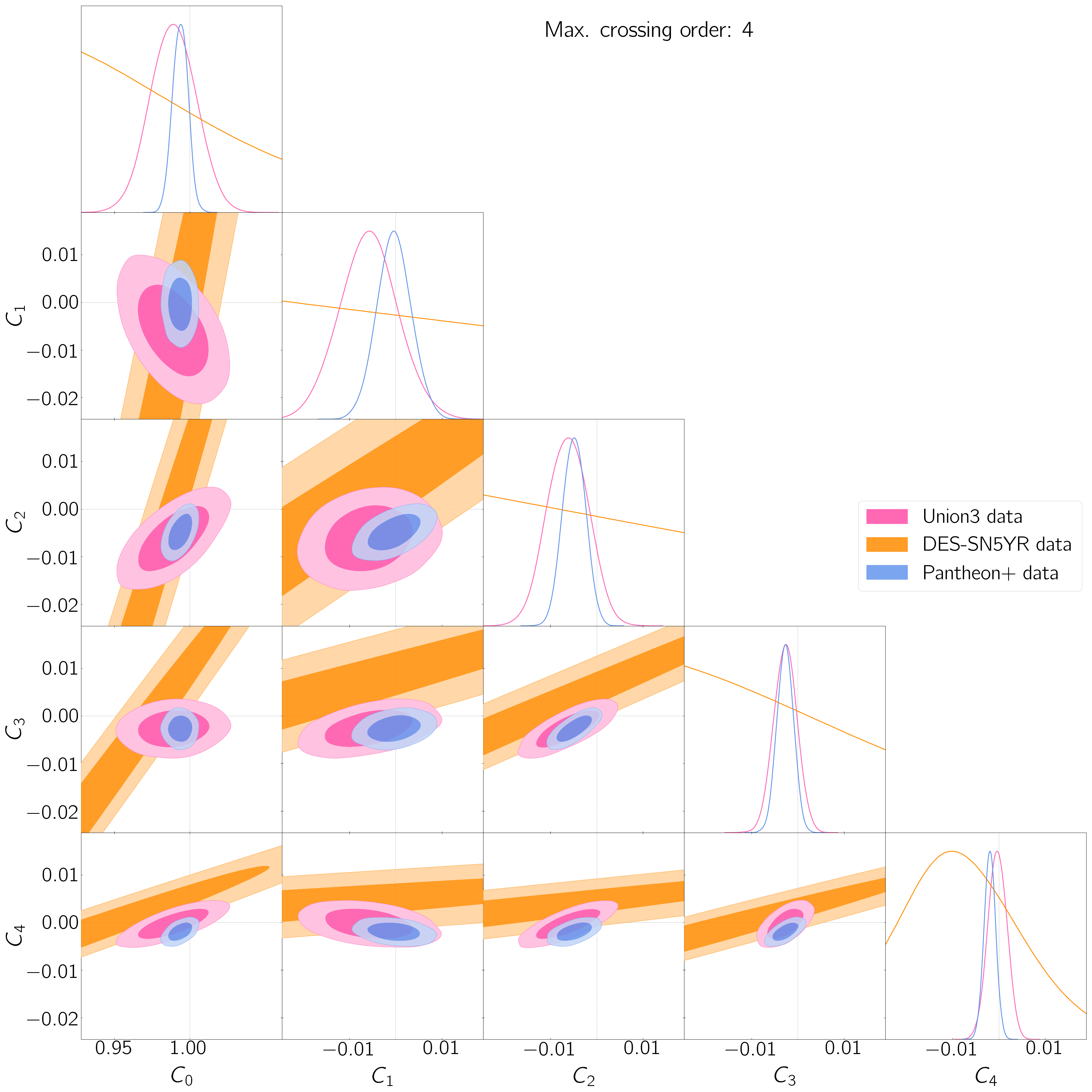}
\caption{Contour plots for the hyperparameters of the \textbf{DES-SN5YR $\Lambda$CDM best fit} template function, deformed to fit to all datasets. The contours from fits to Pantheon+, Union3 and DES-SN5YR data are shown in blue, pink and orange, respectively. We show the results at maximum crossing function order of $3$ (upper panel) and $4$ (lower panel), for the full redshift range.}
\label{D5BFcont34}
\end{figure}

\begin{figure}[H]
\centering
\includegraphics[scale=0.13]{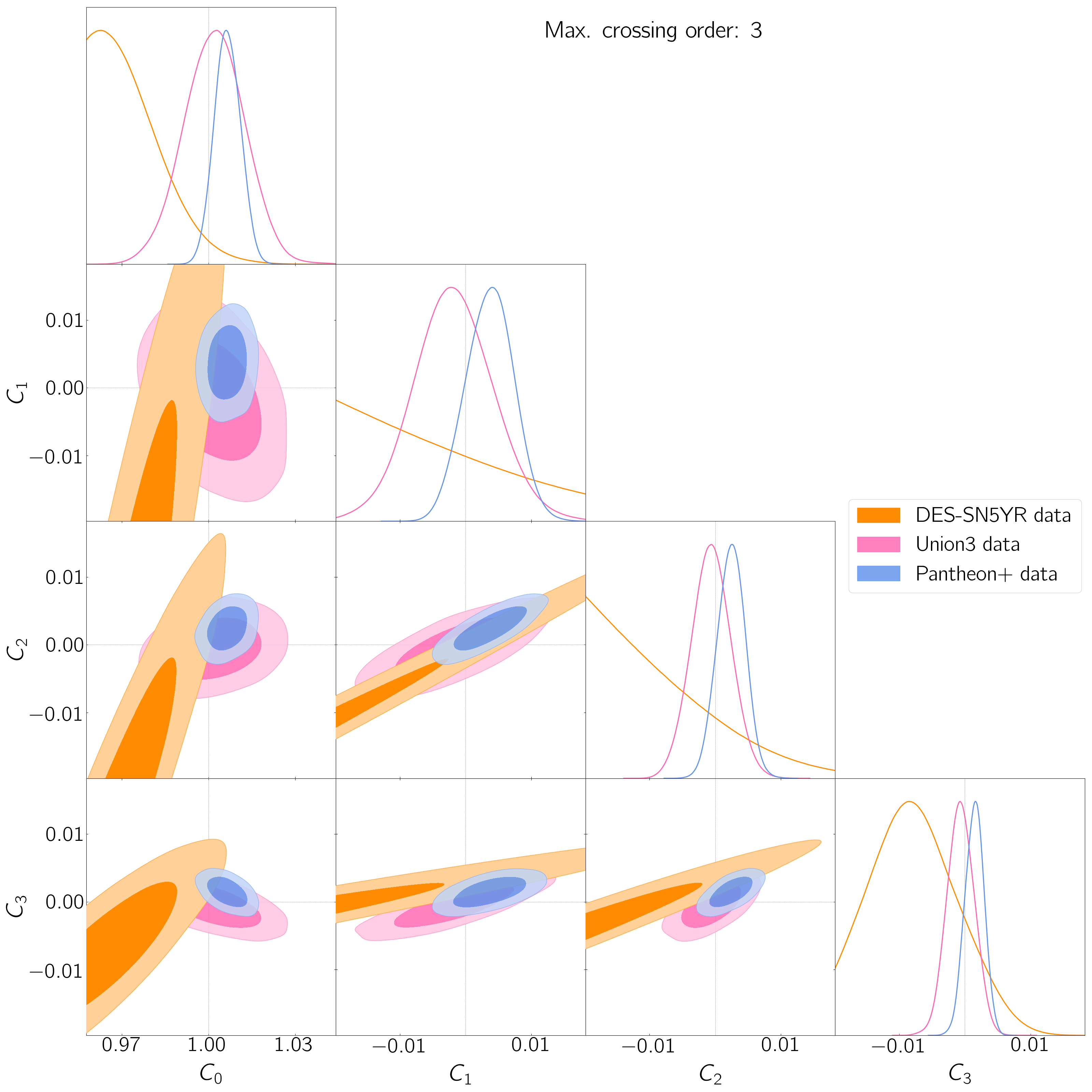}
\includegraphics[scale=0.11]{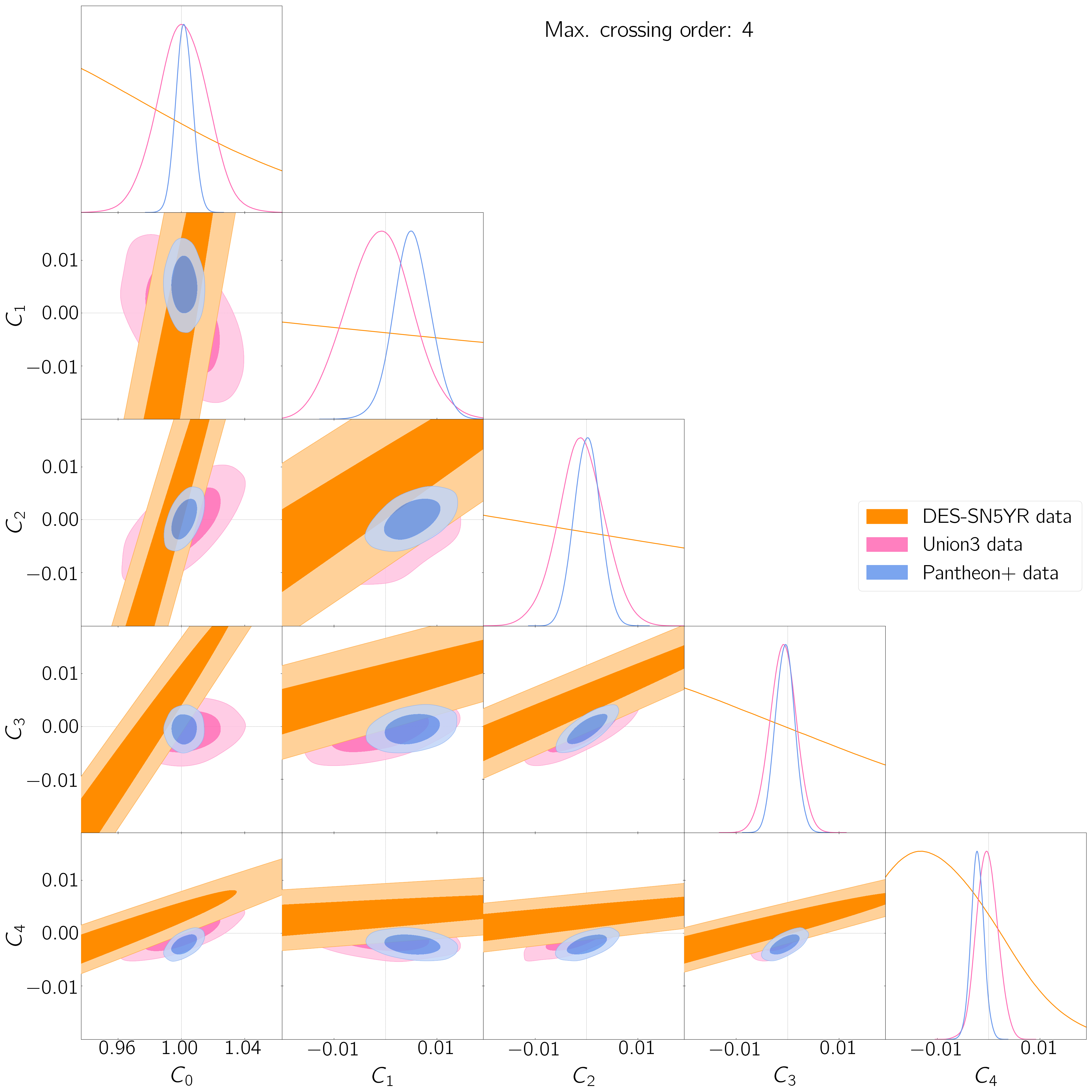}
\caption{Contour plots for the hyperparameters of the \textbf{Union3 iterative smoothing} template function, deformed to fit to all datasets. The contours from fits to Pantheon+, Union3 and DES-SN5YR data are shown in blue, pink and orange, respectively. We show the results at maximum crossing function order of $3$ (upper panel) and $4$ (lower panel), for the full redshift range.}
\label{U3IScont34}
\end{figure}

To avoid the unwieldy proliferation of contour plots, in this appendix we summarise the important information for each part of the analysis. For the sake of space, we abbreviate the dataset to which the deformed functions are fit as ``P+'', ``U3'', ``D5'' for Pantheon+, Union3 and DES-SN5YR respectively, as well as `template function' as ``T.F.'', in the tables that follow. In the case of the initial analysis, including the data over the full range of redshift, we also report the contours plots for the $3^{rd}$ and $4^{th}$ order deformations of the Union3 and DES-SN5YR template functions in the relevant sections (see Figs~\ref{U3BFcont34},~\ref{D5BFcont34} and~\ref{U3IScont34}). 

\subsection{Full redshift ranges}

${ }$\vspace{-20pt}
\begin{table}[H]
\centering
\begin{tabular}{c|c c c|c c c|c c c}
\small
Flat $\Lambda$CDM: &\multicolumn{3}{c|}{Pantheon+}&\multicolumn{3}{c|}{Union3}& \multicolumn{3}{c}{DES-SN5YR} \\
\hline
Data: & P+ & U3  & D5& P+ & U3  & D5& P+ & U3  & D5\\
\hline
$0^{th}$ order&$1$&$\sim2$&$\sim2$&$\sim1$&$\sim1$&$\sim1$&$\sim1$&$\sim1$&$\sim1$\\
$1^{st}$ order&$1$&$2$&$2$&$2$&$1$&$2$&$1$&$1$&$1$\\
$2^{nd}$ order&$1$&$1\sim2$&$ >2  $&$1$&$2$&$ >2  $&$1$&$2$&$  >2 $\\
$3^{rd}$ order&$1$&$1\sim2$&$ >2  $&$1$&$1$&$ >2  $&$1$&$2$&$>2$\\
$4^{th}$ order&$1$&$1$&$1$&$1$&$1$&$1$&$1$&$1$&$1$
\end{tabular}
\caption{$\sigma$-level at which all the crossing hyperparameters at each order are \wlm{compatible} with standard values, for each \textbf{Flat $\Lambda$CDM best fit} template function, fit to each dataset.}
\label{BFcontsum}
\end{table}
${ }$\vspace{-20pt}
\begin{table}[H]
\centering
\begin{tabular}{c|c c c|c c c|c c c}
\small
Iterative Smoothing: &\multicolumn{3}{c|}{Pantheon+}&\multicolumn{3}{c|}{Union3}& \multicolumn{3}{c}{DES-SN5YR} \\
\hline
Data: & P+ & U3  & D5& P+ & U3  & D5& P+ & U3  & D5\\
\hline
$0^{th}$ order&$1$&$>2$&$>2$&$>2$&$1$&$\sim2$&-&-&$1$\\
$1^{st}$ order&$1$&$\sim1$&$2$&$>2$&$1$&$1$&-&-&$1$\\
$2^{nd}$ order&$1$&$2$&$>2$&$1\sim2$&$1$&$>2$&-&-&$1$\\
$3^{rd}$ order&$1$&$2$&$ >2  $&$2$&$1$&$ \sim2  $&-&-&$1$\\
$4^{th}$ order&$1$&$1$&$1$&$1$&$1$&$1$&-&-&$1$
\end{tabular}
\caption{$\sigma$-level at which all the crossing hyperparameters at each order are \wlm{compatible} with standard values, for both \textbf{iterative smoothing} template functions, fit to each dataset.}
\label{IScontsum}
\end{table}

\subsection{Low $z_{max}$ cut}
${ }$\vspace{-20pt}
\begin{table}[H]
\centering
\begin{tabular}{c|c c c|c c c|c c c}
Flat $\Lambda$CDM: T.F. &\multicolumn{3}{c|}{Pantheon+}&\multicolumn{3}{c|}{Union3}& \multicolumn{3}{c}{DES-SN5YR} \\
Data: & P+ & U3  & D5& P+ & U3  & D5& P+ & U3  & D5\\
\hline
$0^{th}$ order&$1$&$\sim2$&$\sim2$&$1$&$1$&$1$&$1$&$1$&$1$\\
$1^{st}$ order&$1$&$2$&$2$&$1$&$2$&$\sim2$&$1$&$2$&$2$\\
$2^{nd}$ order&$1$&$1\sim2$&$ >2  $&$1$&$\sim1$&$ >2  $&$1$&$1\sim2$&$  >2 $
\end{tabular}
\caption{Number of $\sigma$ at which all the crossing hyperparameters at a particular order are \wlm{compatible} with the standard values, for each Flat $\Lambda$CDM best fit template function, fit to each dataset with the low $z_{max}$ cut-off.}
\label{lowzBFcontsum}
\end{table}

${ }$\vspace{-10pt}
${ }$\vspace{-20pt}
\begin{table}[H]
\centering
\begin{tabular}{c|c c c|c c c|c c c}
Iterative Smoothing T.F.: &\multicolumn{3}{c|}{Pantheon+}&\multicolumn{3}{c|}{Union3}& \multicolumn{3}{c}{DES-SN5YR} \\
Data: & P+ & U3  & D5& P+ & U3  & D5& P+ & U3  & D5\\
\hline
$0^{th}$ order&$1$&$\sim2$&$\sim2$&$>2$&$1$&$\sim1$&$\sim2$&$\sim1$&$1$\\
$1^{st}$ order&$1$&$2$&$2$&$>2$&$1$&$1$&$\sim2$&$1$&$1$\\
$2^{nd}$ order&$1$&$1\sim2$&$ >2  $&$1\sim2$&$1$&$ >2  $&$1\sim2$&$1\sim2$&$1 $
\end{tabular}
\caption{Number of $\sigma$ at which all the crossing hyperparameters at a particular order are \wlm{compatible} with the standard values, for each iterative smoothing template function, fit to each dataset with the low $z_{max}$ cut-off.}
\label{lowzIScontsum}
\end{table}
\end{appendices}
${ }$\vspace{-30pt}
\bibliography{main}

\end{document}